\documentclass{aa}

\usepackage[dvips]{graphics} 
\usepackage{epsf} 
\usepackage{amsmath,amssymb} 
\usepackage{latexsym}

\begin{document} 

   \title{The submillimeter C and CO lines in Henize 2-10 and NGC 253.}
   \subtitle{} 
 
 \author{E.Bayet\inst{1}, M.Gerin\inst{1}, T.G.Phillips\inst{2} and A.Contursi\inst{3}
 }

   \offprints{E.Bayet, \email{estelle.bayet@lra.ens.fr}} 
 
   \institute{Laboratoire de Radioastronomie (LRA), Observatoire de Paris and Ecole Normale Sup\'erieure, 24 rue Lhomond, F-75005 Paris, France (CNRS-UMR 8112)
         \and 
           California Institute of Technology, Downs Laboratory of Physics 320-47, Pasadena, CA 91125, USA 
	 \and
	   Max Planck Institute f\"ur Extraterrestrische Physik, Postfach 1312, 85741, Garching, Germany}
 
   \date{Accepted 05/07/04} 
   \abstract{\textbf{The purpose of this paper is to describe a method for
 determining a cooling template for galaxies, using nearby galaxies,
 and applicable to future observations of distant galaxies.}
We observed two starburst galaxies (NGC 253 and Henize 2-10) with
the Caltech Submillimeter Observatory in the rotational lines
   of carbon monoxide $^{12}$CO(J=3-2), (J=6-5) and (J=7-6) for both,
   and also $^{12}$CO(J=4-3) and $^{13}$CO(J=3-2) for Henize 2-10 and in the $^{3}$P$_{2}$-$^{3}$P$_{1}$ fine-structure
   transitions of atomic carbon [CI] at 809 GHz for NGC 253. Some of
   these observations have been made previously, but the present
   multitransition study (including data found in the
   literature) is the most complete to date for the two galaxies.
   From these observations, we have derived the properties of the warm
   and dense molecular gas in the galaxy nuclei. We used an LTE
   analysis and an LVG radiative transfer model to determine physical
   conditions of the interstellar medium in both sources and predicted
 \textbf{integrated line properties} of all CO transitions up to
 $^{12}$CO(15-14). We found the observations to be in good agreement with a
 medium characterized by $T_{k}\approx 50-100\text{ }K$,
 $\frac{^{12}CO}{^{13}CO} \approx 30$, $n(H_{2}) \gtrsim 10^4\text{
 }cm^{-3}$ and $N(^{12}CO)=3.5\pm1.0\times10^{18}\text{ }cm^{-2}$ for
 Henize 2-10 and characterized by $T_{k}\approx 70-150\text{ }K$,
 $\frac{^{12}CO}{^{13}CO} \approx 40$, $n(H_{2}) \gtrsim 10^4\text{
 }cm^{-3}$ and $N(^{12}CO)=1.5 \pm 0.5 \times10^{19}\text{ }cm^{-2}$
 for NGC 253. A PDR model has also been used and here the data are
 well fitted (within 20 \%) by a model cloud with a gas density of
 n(H)= 8.0$\pm 1.0 \times$ 10$^{5}$ $cm^{-3}$ and an incident FUV flux
 of $\chi \approx$20000 for Henize 2-10. For NGC 253, we deduced n(H)=
 3.0$\pm 0.5 \times$ 10$^{5}$ $cm^{-3}$ and $\chi \approx$20000 for
 the modelled cloud. \textbf{The physical properties of warm gas and
 CO cooling curves of the target galaxies are compared
 with those measured for the nucleus of the Milky Way and the Cloverleaf
 QSO. The gas properties and CO cooling curve are similar for the two starburst galaxies and the Cloverleaf QSO while the Milky Way nucleus exhibits lower excitation molecular gas.}\\

 } 
 
\titlerunning{The submillimeter C and CO lines in  Henize 2-10 and NGC 253 } 
\authorrunning{E.Bayet et al.} 
 
   \maketitle 

%________________________________________________________________ 
 
\section{Introduction}\label{sec:intro}
Over the past decade, observations of distant galaxies have become
possible in the millimeter and submillimeter bands and considerable progress
is expected for the next decade. Much of the energy provided by star
formation in those galaxies is down-converted to the submm and appears
as dust emission and the familiar gas cooling lines. Because the
information on distant galaxies is scarce, due to the lack of
resolution and sensitivity, it is essential to understand nearby
galaxies in order to provide templates for more distant objects. In recent
studies of emission of fine-structure transitions of atomic carbon
[CI], ionized atomic carbon [CII] and carbon monoxide (CO), we have
begun to see how they relate to each other and to dust emission (Gerin
\& Phillips 1998, 2000). However, to estimate the relative
contributions of the various species to the gas cooling, it is often
necessary {\bf to make assumptions for the power
 in the various unobserved lines. }

It is well known that the fine-structure lines of ionized carbon [CII]
and atomic oxygen oxygen [OI] contribute most of the cooling of the
neutral interstellar gas in galaxies. These lines trace the cooling of
the diffuse neutral gas with a significant contribution from
Photo-Dissociation Regions (PDRs). Fine-structure lines of ionized
oxygen [OIII] and nitrogen [NII] trace the ionized gas, either in HII
regions([OIII]) or the diffuse ionised gas ([NII]). In molecular gas,
the cooling radiation is due to atomic carbon, carbon monoxide and
water lines (although water (like most molecules) is not sufficiently
widespread to contribute much cooling to a galaxy as a whole). These
predictions have been confirmed by the COBE-FIRAS observations of the
Milky Way: apart from [CII] and [NII] and probably [OI], the most
intense submillimeter lines are from C and CO (see Fixsen et al. 1999
and Sect.~\ref{secsub:cool}). The relative contributions of the
different lines of C and CO vary along the Galactic plane. Also C
contributes less in proportion to the total cooling towards the
Galactic Center than towards the rest of the disk (Bennett et al 1994
and Fixsen et al. 1999). 

The detection of CO lines of distant galaxies remains an extremely
difficult task and will continue so until ALMA is ready. However,
detecting their submillimeter continuum emission is now feasible, with
the advent of sensitive bolometer arrays. Due to the negative
K-correction effect, the dust emission of distant galaxies piles up at
submillimeter wavelengths irrespective of redshift. Additional
information is needed to be able to estimate a photometric redshift,
or to classify the galaxies. For instance, Yun \& Carilli (2002) have
proposed to use the cm radio continuum emission for obtaining
photometric redshifts. From the model developed by Lagache, Dole \&
Puget (2003) to explain the evolution of number counts in the
submillimeter and far infrared (FIR), it appears that two main galaxy
populations contribute to these number counts: a relatively local
population of ``cold'' galaxies, similar to our own galaxy, in which
the continuum emission is dominated by extended disk-type emission,
and an additional, strongly evolving population of infrared (IR)
luminous galaxies , similar to local starburst galaxies. From the
information gathered so far on high-J CO emission in galaxies (this
work; Israel \& Baas 2002, 2003; Bradford et al. 2003; Ward et
al. 2003 and references therein), it is likely that high-J CO lines
($^{12}$CO(4-3) and above) will be more prominent in actively star
forming galaxies compared to the less active galaxies in which only
low-J CO lines and the carbon lines are expected. \textbf{In this
  paper, from an extensive set of observations of submillimeter CO
  lines, we conclude that, relative to $^{12}$CO(1-0), starburst galaxies do exhibit stronger high-excitation CO lines ($^{12}$CO(4-3) and above), than the Milky Way galaxy.}

\textbf{Determining the physical conditions of molecular clouds in external
galaxies can be difficult. For single-dish observations of external
galaxies, the beam size is usually larger than the angular diameter of
typical molecular clouds, hence the cloud emission is beam-diluted. In
addition, molecular clouds are known to be clumpy (e.g. Stutzki \&
Guesten 1990). 
The intensities of CO rotational lines are sensitive to the
  gas physical conditions (H$_2$ density n(H$_2$), and kinetic
  temperature $T_k$),  the $^{12}$CO column density N($^{12}$CO) as well as the 
velocity field. In this paper, we use several methods for determining
the $^{12}$CO column density and the physical conditions, and for predicting
the intensity of all CO rotational lines up to $^{12}$CO(15-14). The LTE
  analysis is useful for constraining the kinetic temperature. Using
LVG models, three parameters (n(H$_2$),$T_k$, N($^{12}$CO)) can be measured. PDR
  models include a self consistent treatment of the thermal and
  chemical processes and provide therefore a better theoretical
  understanding of the gas properties.}

\begin{figure}[h!]
\begin{center}
      \epsfxsize=9cm 
      \epsfbox{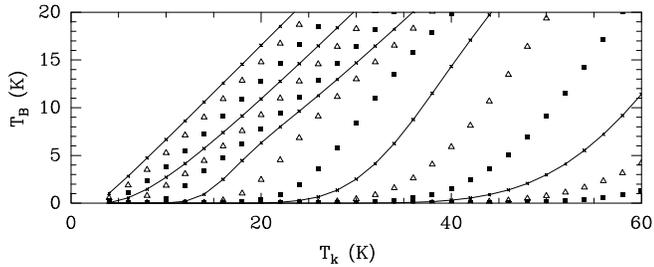} 
    \caption{\textbf{Brightness temperature $T_{B}$ (equations ~\ref{Tas}
    and ~\ref{Jnu} in Sect.~\ref{sec:intro}) vs. the kinetic
    temperature, $T_{k}$, assuming a unity filling factor ($FF = 1$). 
The first curve on the left is $^{12}$CO(1-0) (full line) followed by $^{12}$CO(2-1) (triangles), $^{12}$CO(3-2) (squares), $^{12}$CO(4-3) (full line)... up to $^{12}$CO(15-14). For this plot, we used N(CO)= $1\times 10^{18}$ $cm^{-2}$ and $\Delta v= 1$ $kms^{-1}$ resulting in $\tau >> 1$.\label{fig:planck}}}
\end{center}
\end{figure}

We recall that, assuming LTE, the brightness temperature, $T_{B}$, of CO lines as a function of the kinetic temperature $T_{k}$ can be written as:
\vspace*{-0.5cm}
\begin{center}
\begin{equation}\label{Tas}
T_{B} = FF\left(J_{\nu}(T_{k})-J_{\nu}(T_{bg})\right)\left(1 - e^{-\tau}\right) 
\end{equation}
\end{center}
where $FF$ is the source surface filling factor in the beam, $\tau$ 
is the line opacity, $T_{bg}$ is the CMB temperature ($T_{bg}=2.7$
$K$) and $J_{\nu}(T_{k})$ is described in Kutner \& Ulich 1981 by:
\vspace*{-0.5cm}
\begin{center}
\begin{equation}\label{Jnu}
J_{\nu}(T_{k}) = \left(\frac{h\nu}{k_{b}}\right)\times\frac{1}{exp(\frac{h\nu}{k_{b}T_{k}}) -1}
\end{equation}
\end{center}
where $k_{b}$ is the Boltzmann's constant and 
$h$ is the Planck constant. We show in Fig.~\ref{fig:planck} the
brightness temperature, $T_{B}$, as a function of the kinetic
temperature $T_{k}$. In this example, typical figures for Galactic
molecular clouds have been chosen with $FF$ = 1, $N(^{12}CO)=1\times10^{18}\text{ }cm^{-2}$ and $\Delta v= 1$
$kms^{-1}$, resulting in saturated CO lines.

For a gas temperature around $T_{k} \approx 10\text{ }K$, low-J CO
lines ($^{12}$CO(1-0), $^{12}$CO(2-1) and $^{12}$CO(3-2)) are easily
detected while mid and high-J CO lines ($^{12}$CO(5-4), $^{12}$CO(6-5)
and above) need warmer gas to be detected ($T_{k} \gtrsim 15\text{
}K$).

Since we want to sample warm gas, we have observed, for both galaxies,
the high-J CO lines: $^{12}$CO(4-3) (only for Henize
2-10), $^{12}$CO(6-5) and $^{12}$CO(7-6) which are more sensitive
to warm, dense gas directly involved in the starburst. We also
observed $^{12}$CO(3-2) for both galaxies and $^{13}$CO(3-2) for
Henize 2-10, to better constrain the models (see
Sect.~\ref{secsub:lte}).

 Atomic carbon has also proved to be a good tracer of molecular gas (Israel \& Baas 2001 and Gerin \& Phillips 2000). The fine structure transition of atomic carbon $^{3}$P$_{2}$-$^{3}$P$_{1}$[CI] at 809 GHz for NGC 253 has also
been observed here, since the intensity ratio between this line and
the $^{3}$P$_{1}$-$^{3}$P$_{0}$[CI] at 492 GHz is a sensitive tracer
of the total gas pressure. For NGC 253, Bradford et al. (2003) and
Israel, White \& Baas (1995) give the intensity of the fine structure
transition of atomic carbon $^{3}$P$_{1}$-$^{3}$P$_{0}$[CI] at 492
GHz. \textbf{For Henize 2-10, Gerin \& Phillips (2000) give the intensity of
the $^{3}$P$_{1}$-$^{3}$P$_{0}$[CI] at 492 GHz but we could not find a
measure of $^{3}$P$_{2}$-$^{3}$P$_{1}$[CI] at 809 GHz to compute the
line ratio.}

In the following, we present new CO and CI data for NGC 253 and for
Henize 2-10 (see Sect.~\ref{sec:obs}). The galaxies are described in
Sec.~\ref{sec:sample}, the observations in Sec.~\ref{sec:obs}. \textbf{The
results and data analysis providing ISM properties are discussed in
Sec.~\ref{sec:data} and in Sec.~\ref{sec:mod}, respectively, with conclusions in section 6.}

%______________________________________________________________________________

\section{The sample:}\label{sec:sample}
Henize 2-10 is a blue compact dwarf galaxy with a metallicity of 12+log($\frac{O}{H})\approx 8.93$ (Zaritsky, Kennicutt \& Huchra 1994). It presents a double core structure suggesting a galaxy merger (Baas, Israel \& Koornneef 1994). From optical, infrared and radio observations of Henize 2-10, the derived distance is 6 Mpc (Johansson 1987). Henize 2-10 is rich in neutral atomic hydrogen (Allen, Wright \& Goss 1976) and harbors a considerable population of Wolf-Rayet stars (D'Odorico, Rosa \& Wampler 1983; Kawara et al. 1987). Previous CO observations have been reported by Baas, Israel \& Koornneef (1994); Kobulnicky et al. (1995 and 1999) and Meier et al. (2001). The starburst is confined in the galaxy nucleus within a radius of 5''(150 pc). This galaxy was chosen because it shows intense and narrow CO lines.\\
NGC 253 is a highly inclined galaxy (i=78$^{o}$, Pence 1980) of type Sc. With
M82, it is the best nearby example of a nuclear starburst (Rieke, Lebofsky \& Walker
1988). Blecha (1986) observed 24 of the brightest globular clusters and estimated a distance of 2.4 to 3.4 Mpc, while Davidge \& Pritchet (1990) analysed a color-magnitude diagram of stars in the halo of NGC 253 and conclude that the distance is 1.7 to 2.6 Mpc. In the following, we shall use a value of D = 2.5 Mpc as in Mauersberger et al. (1996). NGC 253 has been extensively
observed at all wavelengths (Turner 1985 (map in all four 18 cm lines of OH in the nucleus); Antonucci \& Ulvestad 1988 (deep VLA maps of NGC 253 which show at least 35 compact radio sources); Carlstrom et al. 1990 (the J = 1-0 transitions of HCN and HCO$^{+}$ and the 3 mm continuum emission); Telesco \& Harper 1980 (30 and 300 $\mu$m emission); Strickland et al. 2004 (X-ray observations)). NGC 253 presents a very active and young starburst. It is among the first galaxies to have been detected in submillimeter lines and continuum. The NGC 253 nucleus has been
mapped previously in various lines of CO and C (Bradford et al. 2003;
Israel \& Baas 2002; Sorai et al. 2000; Harrison, Henkel \& Russell
1999; Israel, White \& Baas 1995; Wall et al. 1991; Harris et al. 1991). The most intense emission is associated with the starburst nucleus with additional weak and extended emission from the disk (Sorai et al 2000).
Both galaxies harbor superstar clusters in their nuclei (Keto et
al. 1999) similar to the superstar clusters found in interacting
galaxies (Mirabel et al. 1998). Basic properties for both galaxies are
summarized in Table~\ref{tab:prop}.
 
\begin{table}
\caption{Basic properties of Henize 2-10 and NGC 253.}
         \label{tab:prop}
	 $$ 
	 \begin{array}{p{0.5\linewidth}l}
	   \hline
	   \noalign{\smallskip}
	   HENIZE 2-10&\\
	   \noalign{\smallskip}
	   \hline
	   \noalign{\smallskip}
	   RA(1950) & 08:34:07.2\\
	   DEC(1950) & -14:26:06.0\\
	   Distance & 6 Mpc^{1}\\
	   Velocity (LSR) & +850\text{ }kms^{-1}\\
	   Optical size & 30'' \times 40 ''\\
	   Position Angle & 130^{o}\\
	   Inclination & 44^{o}\\
	   Metallicity & 12+log(\frac{O}{H})\approx 8.93^{2}\\
	   L$_{FIR}$ & 1.7\times10^{9}\text{ }L_{\odot}\text{ }^{3}\\
	   Absolute Luminosity L$_{B}$ & 1.2\times10^{9}\text{ }L_{\odot}\text{ }^{4}\\
	   \noalign{\smallskip}
	   \hline
	   \hline
	   \noalign{\smallskip}
	   NGC 253&\\
	   \noalign{\smallskip}
	   \hline
	   \noalign{\smallskip}
	   RA(1950) & 00:45:05.7\\
	   DEC(1950) & -25:33:38.0\\
	   Distance & 2.5 Mpc^{5}\\
	   Velocity (LSR) & +240\text{ }kms^{-1}\\
	   Optical size & 27.5' \times 6.8'\\
	   Position Angle & 51^{o}\\
	   Inclination & 78.5^{o}\\
	   Metallicity & 12+log(\frac{O}{H})= 8.99\pm 0.31^{2}\\
	   L$_{FIR}$ & 1.0\times10^{10}\text{ }L_{\odot}\text{ }^{3}\\
	   Absolute Luminosity L$_{B}$ & 2.7\times10^{10}\text{ }L_{\odot}\text{ }^{6}\\
	   \noalign{\smallskip}
	   \hline
	 \end{array}
	 $$
$^{1}$: Johansson 1987\\
$^{2}$: Zaritsky et al. 1994\\
$^{3}$: Computed from formula in Imanishi \& Dudley (2000) and from Sanders \& Mirabel (1996)\\
$^{4}$: Baas, Israel \& Koornneef 1994\\
$^{5}$: As in Mauersberger et al. 1996 (see text)\\
$^{6}$: Pence 1980\\
\end{table} 

%___________________________________________________________________________

\section{Observations}\label{sec:obs}  
\begin{figure}
\vspace*{1cm}
\begin{center}
    \epsfxsize=9cm 
    \epsfbox{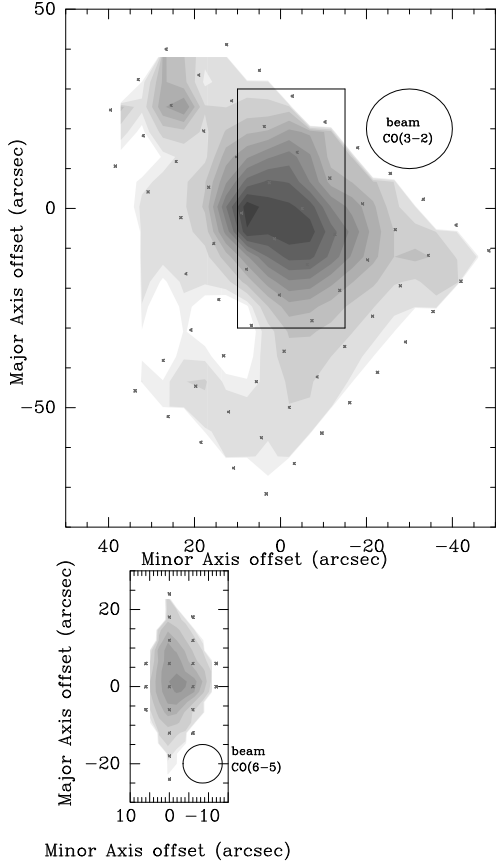} 
  \caption{Map of the velocity integrated intensity of $^{12}$CO(3-2)
    (Top) and $^{12}$CO(6-5) (Bottom) towards NGC 253 nucleus. The CO
    emission is integrated over the full velocity range -~100 to 700
    $kms^{-1}$ and 60 to 400 $kms^{-1}$, respectively. For both maps,
    missing grid positions are interpolated and the crosses show the
    observed positions. The first and the second contours of the
    $^{12}$CO(3-2) map are of 41.6 $Kkms^{-1}$ and 81.6 $Kkms^{-1}$
    respectively; others contours range from $\int{T_{mb}dv}$ = 81.6
    to 816 $Kkms^{-1}$ in steps of 81.6 $Kkms^{-1}$. For the
    $^{12}$CO(6-5) map, contours are of 43 $Kkms^{-1}$, 141
    $Kkms^{-1}$, 282 $Kkms^{-1}$, 564 $Kkms^{-1}$, 846 $Kkms^{-1}$,
    1128 $Kkms^{-1}$ and 1410 $Kkms^{-1}$. The peak value of the
    $^{12}$CO(3-2) map is $\int{T_{mb}dv} \approx 816$ $Kkms^{-1}$. The
    peak value of the $^{12}$CO(6-5) map is $\int{T_{mb}dv} \approx
    1400$ $Kkms^{-1}$. The maps are rotated by 51$^{0}$ to match the
    major axis position angle of the CO emission in NGC 253.\textbf{
      The $^{12}$CO(3-2) map has been shifted by -5'' in minor axis
      offset (see text in Sect.~\ref{sec:obs} and in
      Sect.~\ref{secsub:speketmap}). The black box in the upper figure 
represents the size of the lower figure}.}\label{fig:map}  
\end{center}
\end{figure}

The observations were made during various sessions at the Caltech Submillimeter
Observatory (CSO) in Hawaii (USA) with the Superconducting Tunnel
Junction receivers operated in double-side band mode. The atmospheric
conditions varied from good ($\tau_{225} \lesssim 0.1$) to excellent
($\tau_{225}\approx$ 0.06). We used a chopping secondary mirror with a
throw of 1 to 3 arcmin, depending on the size of the source, and with
a frequency around 1 Hz. Henize 2-10 is point-like for all lines. The NGC 253 nucleus has a size of 50'' (see Fig.~\ref{fig:map}) which
is smaller than the chopping throw. For this galaxy, we used a 3' chopping
throw in $^{12}$CO(3-2). There is no sign of contamination by
emission in the off beams. We restricted the chopping throw to 1' for
$^{12}$CO(6-5) and $^{12}$CO(7-6) as the emission is very compact in
these lines. Spectra were measured with two acousto-optic
spectrometers (effective bandwidth of 1000 MHz and 500 MHz). The first
one has a spectral resolution about 1.5 MHz and the second one about 2
MHz. The IF frequency of the CSO receivers is 1.5 GHz. The main beam
efficiencies ($\eta$) of the CSO were 69.8\%, 74.6\%, 51.5\%, 28\% and
28\% at 230, 345, 460, 691 and 806 GHz respectively, as measured on
planets. We used the ratio $\frac{1}{\eta}$ to convert T$^{*} _{A}$
into T$_{mb}$. The beam size at 230, 345, 460, 691 and 806 GHz is
30.5'', 21.9'', 14.5'', 10.6'' and 8.95'' respectively \footnote{See
  web site: http://www.submm.caltech.edu/cso/}.

The pointing was checked using planets (Jupiter, Mars and Saturn) and
evolved stars (IRC 10216 and R-Hya). The pointing accuracy is around
5''. The overall calibration accuracy is 20\%.
Data have been reduced using the CLASS data analysis package. 
The spectra have been smoothed to a velocity resolution of 10
kms$^{-1}$ and linear baselines have been removed.

For NGC 253, we obtained spectra towards the central position for both
$^{12}$CO(7-6) line and the atomic carbon
CI($^{3}$P$_{2}$-$^{3}$P$_{1}$) line simultaneously. We observed 70
positions in the $^{12}$CO(3-2) line and 20 positions for the
$^{12}$CO(6-5) line. For Henize 2-10 we observed only the nucleus for the $^{12}$CO(3-2), $^{12}$CO(4-3), $^{12}$CO(6-5),
$^{12}$CO(7-6) and $^{13}$CO(3-2) lines. The carbon line
$^{3}$P$_{1}$-$^{3}$P$_{0}$ [CI] is from Gerin \& Phillips (2000).

\begin{figure}
\begin{center}
      \epsfxsize=9cm 
      \epsfbox{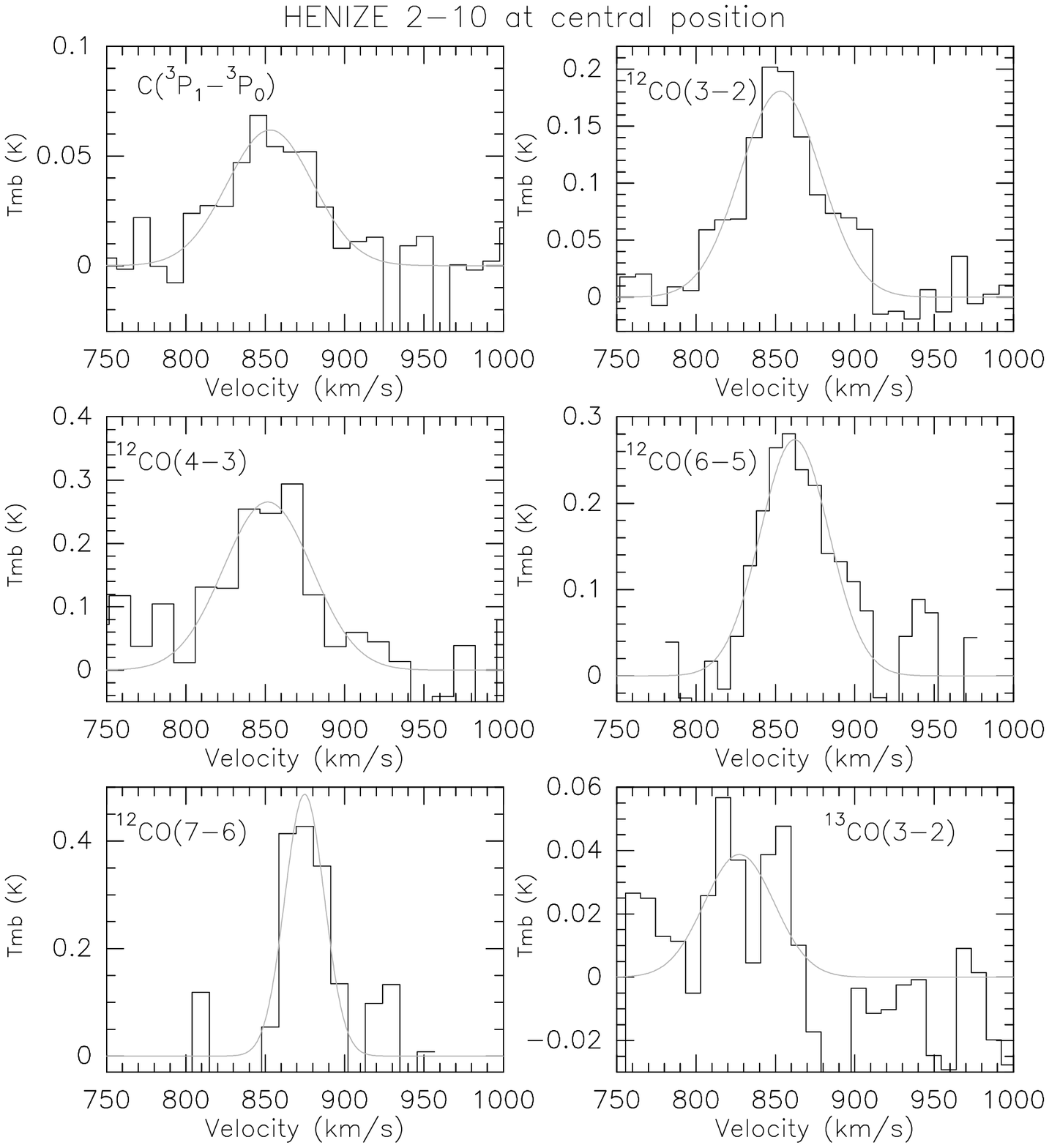} 
      \epsfxsize=8cm 
      \epsfbox{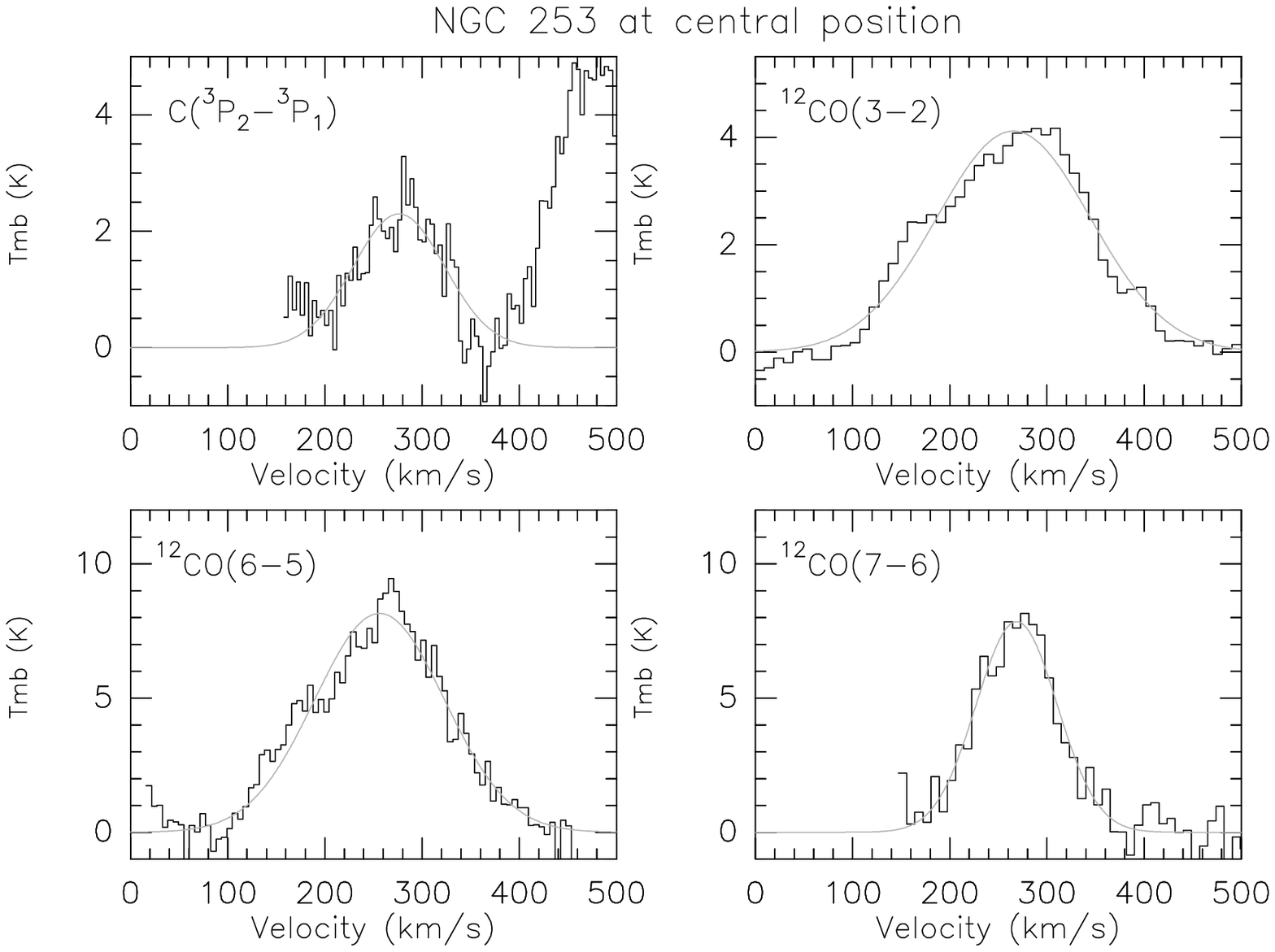}
    \caption{Spectra of Henize 2-10 \textbf{(top six)} and NGC 253 \textbf{(bottom four) }at central
    positions as given in Table~\ref{tab:prop}. \textbf{Velocities (horizontal axis)
    are given relative to the LSR (V$_{LSR}$) in $kms^{-1}$ and the line
    intensities (vertical axis) are in units of T$_{mb}$ ($K$). The
    grey curves are Gaussian fits. For NGC 253, the line on the right hand side of CI($^{3}$P$_{2}$-$^{3}$P$_{1}$) is the edge of $^{12}$CO(7-6) line seen in the image side band.}}\label{fig:spec}
\end{center}
\end{figure}

\begin{table*}
  \caption{Results for Henize 2-10.}\label{tab:obs1}
  \begin{center}
    \begin{tabular}{|c|c|c|c|c|c|c|}
      \hline
      \textbf{HENIZE 2-10}&&&&&&\\
      \hline
      Transition & Freq & Beam Size & $\int(T_{mb}dv)$ & Intensity & Flux & References$^{a}$\\
      &(GHz) & ('') & (Kkms$^{-1}$) & (Wm$^{-2}$sr$^{-1}$) & (Wm$^{-2}$) &\\
      \hline
      CI($^{3}$P$_{1}$-$^{3}$P$_{0}$) & 492.162 &14.55 & 4.2$\pm$0.8 & 5.1$\times10^{-10}$ & 3.0$\times10^{-18}$ & 4 \& 5$^{b}$\\
      & & 21.90 & 2.5$\pm$0.5 & 3.0$\times10^{-10}$ & 3.8$\times10^{-18}$ & 4 \& 5$^{b}$\\ 
      \hline
      $^{12}$CO(1-0) & 115.271 & 40.00 & 10.0$\pm$0.8 & 1.6$\times10^{-11}$ & 6.7$\times10^{-19}$ & 1\\
      & & 21.90 & 27.3$\pm$2.2 & 4.3$\times10^{-11}$ & 5.5$\times10^{-19}$ & 1\\ 
      &  & 55.00 & 4.9$\pm$0.2 & 7.6$\times10^{-12}$& 6.1$\times10^{-19}$ & 2\\
      & & 21.90 & 23.9$\pm$1.0& 3.7$\times10^{-11}$ &4.8$\times10^{-19}$ & 2*\\
      $^{12}$CO(2-1) & 230.538 & 21.00 & 17.3$\pm$1.4 & 2.2$\times10^{-10}$ & 2.5$\times10^{-18}$ & 1*\\
      &  & 27.00 & 6.8$\pm$0.8 & 8.5$\times10^{-11}$ & 1.7$\times10^{-18}$ & 2\\
      & & 21.90 & 9.4$\pm$1.1 & 1.2$\times10^{-10}$ & 1.5$\times10^{-18}$ & 2\\
      $^{12}$CO(3-2) & 345.796 & 21.90 & 11.5$\pm$2.3 & 4.9$\times10^{-10}$ & 6.2$\times10^{-18}$ &5* \\
      & & 21.00 & 23.2$\pm$2.1 & 9.8$\times10^{-10}$ & 1.2$\times10^{-17}$ & 1\\
      &  & 22.00 & 16.6$\pm$0.6 & 7.0$\times10^{-10}$ & 9.0$\times10^{-18}$ & 3\\
      $^{12}$CO(4-3) & 461.041 & 14.55 & 18.6$\pm$4.2 & 1.9$\times10^{-9}$ & 1.1$\times10^{-17}$ &5 \\
      & & 21.90 & 10.9$\pm$2.4 & 1.1$\times10^{-9}$ & 1.4$\times10^{-17}$ & 5*\\
      $^{12}$CO(6-5) & 691.473 & 10.60 & 15.7$\pm$3.1 & 5.3$\times10^{-9}$ & 1.6$\times10^{-17}$ & 5\\
      & & 21.90 & 6.8$\pm$1.3 & 2.3$\times10^{-9}$ & 2.9$\times10^{-17}$ & 5*\\
      $^{12}$CO(7-6) & 806.652 & 8.95 & 15.2$\pm$3.0 & 8.2$\times10^{-9}$ & 1.7$\times10^{-17}$ & 5\\
      &  & 21.90 & 5.8$\pm$1.2 & 3.1$\times10^{-9}$ & 4.0$\times10^{-17}$ & 5*\\
      \hline
      $^{13}$CO(1-0) & 110.201 & 40.00 & $<$ 0.5 & $<$ 6.8$\times10^{-13}$& $<$ 2.9$\times10^{-20}$ & 1 \\
      &  & 21.90 & $<$ 1.4 & $<$ 1.8$\times10^{-12}$ & $<$ 2.4$\times10^{-20}$ & 1\\
      &  & 57.00 & 0.3$\pm$0.1 & 3.9$\times10^{-13}$ & 3.4$\times10^{-20}$ & 2\\
      &  & 21.90 & 1.6$\pm$0.5 & 2.2$\times10^{-12}$ & 2.8$\times10^{-20}$ & 2*\\
      $^{13}$CO(2-1) & 220.399 & 21.00 & 0.9$\pm$0.2 & 9.9$\times10^{-12}$ & 1.2$\times10^{-19}$ & 1*\\
      $^{13}$CO(3-2) & 330.588 & 14.00 & 2.3$\pm$0.6& 8.5$\times10^{-11}$ & 4.4$\times10^{-19}$ & 1\\
      &  & 21.90 & 1.3$\pm$0.3 & 4.8$\times10^{-11}$&6.1$\times10^{-19}$ & 1*\\
      &  & 21.90 & 1.8$\pm$0.5 & 6.7$\times10^{-11}$&8.5$\times10^{-19}$ & 5\\
      \hline
      \hline
    \end{tabular}
  \end{center}
\textbf{$^{a}$ References: 1: Baas, Israel \& Koornneef 1994; 2: Kobulnicky et al. 1995; 3: Meier et al. 2001; 4: Gerin \& Phillips 2000; 5: this work; *: used for models.}
\textbf{$^{b}$ We analysed again spectra from Gerin \& Phillips 2000.}
\end{table*}

\begin{table*}
  \caption{Results for the nucleus of NGC 253.}\label{tab:obs2}
  \begin{center}
    \begin{tabular}{|c|c|c|c|c|c|c|}
      \hline
      \textbf{NGC253}&&&&&&\\
      \hline
      Transition & Freq & Beam Size & $\int(T_{mb}dv)$ & Intensity & Flux & References$^{a}$\\
      &(GHz) & ('') & (Kkms$^{-1}$) & (Wm$^{-2}$sr$^{-1}$) & (Wm$^{-2}$) &\\
      \hline
      CI($^{3}$P$_{1}$-$^{3}$P$_{0}$) & 492.162 & 10.20 & 575.0$\pm$115.0$^{b}$ & 7.0$\times10^{-8}$ & 1.9$\times10^{-16}$ & 5\\
      &  & 21.90 & 204.0$\pm$41.0 & 2.5$\times10^{-8}$ & 3.2$\times10^{-16}$ & 5\\
      &  & 43.00 & 98.0$\pm$19.6$^{b}$& 1.2$\times10^{-8}$ & 5.9$\times10^{-16}$ & 5\\
%      &  & 10.20 & 486.0$\pm$60 & 5.9$\times10^{-8}$ & 1.6$\times10^{-16}$ & \\
%      &  & 21.90 & 173.0 & 2.1$\times10^{-8}$ & 2.7$\times10^{-16}$ & \\
      &  & 22.00 & 290.0$\pm$45 & 3.5$\times10^{-8}$ & 4.6$\times10^{-16}$ & 10*\\
      &  & 23.00 & 320.0$\pm$64.0$^{b}$& 3.9$\times10^{-8}$ & 5.5$\times10^{-16}$ & 4\\
      CI($^{3}$P$_{2}$-$^{3}$P$_{1}$) & 809.902 & 8.95 & 188.5$\pm$37.7 & 1.0$\times10^{-7}$ & 2.2$\times10^{-16}$ & 11\\
      &  & 21.9 & 58.1$\pm$11.6 & 3.2$\times10^{-8}$ & 4.0$\times10^{-16}$ & 11*\\
      \hline
      $^{12}$CO(1-0) & 115.271 & 43.00 & 343.0$\pm$68.6$^{b}$& 5.4$\times10^{-10}$ & 2.6$\times10^{-17}$ & 2\\
      &  & 23.00 & 920.0$\pm$82.8 & 1.4$\times10^{-9}$& 2.0$\times10^{-17}$ & 6*\\
      $^{12}$CO(2-1) & 230.538 & 23.00 & 1062.0$\pm$116.8 & 1.3$\times10^{-8}$ & 1.9$\times10^{-16}$ & 7*\\
      &  & 21.00 & 926.0$\pm$185.2$^{b}$& 1.2$\times10^{-8}$ &1.4$\times10^{-16}$ & 2\\
      $^{12}$CO(3-2) & 345.796 & 21.90 & 815.6$\pm$163.1 & 3.5$\times10^{-8}$ & 4.4$\times10^{-16}$ & 11*\\
      &  & 23.00 & 998.0$\pm$139.7 & 4.2$\times10^{-8}$ & 5.9$\times10^{-16}$ & 7\\
      &  & 23.00 & 1194.0$\pm$238.8$^{b}$& 5.0$\times10^{-8}$ & 7.1$\times10^{-16}$ & 2\\
      &  & 14.00 & 1200.0$\pm$240.0$^{b}$& 5.1$\times10^{-8}$ & 2.6$\times10^{-16}$ & 5\\
      &  & 21.90 & 618.9$\pm$123.8 & 2.6$\times10^{-8}$ & 3.3$\times10^{-16}$ & 5\\
      &  & 22.00 & 680.0$\pm$60.0 & 2.9$\times10^{-8}$ & 3.7$\times10^{-16}$ & 8\\
      $^{12}$CO(4-3) & 461.041 & 15.00 & 507.0$\pm$101.4$^{b}$ & 5.1$\times10^{-8}$ & 3.0$\times10^{-16}$ & 3\\
      &  & 21.90 & 285.3$\pm$57.1 & 2.9$\times10^{-8}$ & 3.7$\times10^{-16}$ & 3\\
      &  & 22.00 & 1019.0$\pm$120 & 1.0$\times10^{-7}$ & 1.3$\times10^{-15}$ & 9*\\
      &  & 10.40 & 2160.0$\pm$432.0$^{b}$ & 2.1$\times10^{-7}$ & 6.2$\times10^{-16}$ & 5\\
      &  & 21.90 & 786.2$\pm$157.2 & 7.9$\times10^{-7}$ & 1.0$\times10^{-15}$ & 5\\
%     &  & 43.00 & 285$\pm$? & 2.9$\times10^{-8}$ & 1.4$\times10^{-15}$ & 3\\
      $^{12}$CO(6-5) & 691.473 & 10.60 & 1394$\pm$278.8 & 4.7$\times10^{-7}$ & 1.4$\times10^{-15}$ & 11\\
      &  & 21.90 & 518.3$\pm$104.3 & 1.8$\times10^{-7}$ & 2.2$\times10^{-15}$ & 11*\\
      &  & 8/30 & 861$\pm$258.3 & 2.9$\times10^{-7}$ & 5.0$\times10^{-15}$ & 1\\
      $^{12}$CO(7-6) & 806.652 & 8.95 & 810.2$\pm$162.0 & 4.3$\times10^{-7}$ & 9.3$\times10^{-16}$ & 11\\
      &  & 21.90 & 249.9$\pm$50.3 & 1.3$\times10^{-7}$ & 1.7$\times10^{-15}$ & 11*\\
      &  & 11.5/60 & 1370$\pm$411 & 7.3$\times10^{-7}$ & 2.6$\times10^{-15}$ & 10\\
      \hline
      $^{13}$CO(1-0) & 110.201 & 23.00 & 80$\pm$8.0 & 1.1$\times10^{-10}$&1.5$\times10^{-18}$ & 7*\\
      $^{13}$CO(2-1) & 220.399 & 23.00 & 82$\pm$9.8 & 9.0$\times10^{-10}$ & 1.3$\times10^{-17}$ & 7*\\
      &  & 21.00 & 104$\pm$20.8$^{b}$& 1.1$\times10^{-9}$ & 1.3$\times10^{-17}$ & 2\\
      $^{13}$CO(3-2) & 330.588 & 23.00 & 90$\pm$12.6 & 3.3$\times10^{-9}$ & 4.7$\times10^{-17}$ & 7*\\
      &  & 23.00 & 210$\pm$42.0$^{b}$ & 7.8$\times10^{-9}$ &1.1$\times10^{-16}$ & 2\\
      \hline
    \end{tabular}
  \end{center}
\textbf{$^{a}$ References: 1: Harris et al. 1991; 2: Wall et al. 1991; 3: Guesten et al. 1993; 4: Harrison et al. 1995; 5: Israel, White \& Baas 1995; 6: Mauersberger et al. 1996; 7: Harrison, Henkel \& Russell 1999; 8: Dumke et al. 2001; 9: Israel \& Baas 2002; 10: Bradford et al. 2003; 11: this work; *: used for models.\\$^{b}$ The error is not given in the referenced paper. It has been estimated to be 20\%.}
\end{table*}

%_____________________________________________________________________________

\section{Data analysis}\label{sec:data} 

%________________________________
\subsection{Spectra and maps:}\label{secsub:speketmap}
Henize 2-10 and NGC 253 spectra at central positions are shown in
Fig.~\ref{fig:spec}; Fig.~\ref{fig:map} shows $^{12}$CO(3-2) and
$^{12}$CO(6-5) maps of NGC 253. We detected $^{12}$CO(3-2),
$^{12}$CO(6-5) and $^{12}$CO(7-6) in both galaxies. For these positions, line intensities (A in $Kkms^{-1}$ and I in
$Wm^{-2}sr^{-1}$) and line fluxes (F in $Wm^{-2}$) resulting from
Gaussian fits are listed in Table~\ref{tab:obs1} and in
Table~\ref{tab:obs2}, together with the beam size of the
observations. \textbf{Because of likely pointing offsets ($\approx$
  5'' from pointing scans on planets and evolved stars) between
  different observing runs, we believe that differences in the peak
  positions of $^{12}$CO(3-2) and $^{12}$CO(6-5) in NGC 253 are partly due to pointing errors. But we can not exclude
  that a difference in peak positions remains after the pointing
  correction (see Fig.~\ref{fig:map}). Three peaks appear on the high
  resolution $^{12}$CO(1-0) maps obtained with the Nobeyama
  Interferometer (NRO), one near the center and two on each side at
  around 3 '' south-west and 8 '' north-east respectively (Paglione et
  al, 2004). The offset peak in the $^{12}$CO(3-2) map may be
  associated with the north-east secondary peak.} Table~\ref{tab:obs1} and Table~\ref{tab:obs2} also report relevant data found in the literature. For some spectra, a Gaussian may not represent the true line profile but given the weakness of the signal, fitting anything more complicated is not warranted.\\
To be able to compare intensities of the different CO lines, we have also computed the line intensities (A and I) and fluxes (F) for a common beam size of 21.9'', which is the size of the $^{12}$CO(3-2) beam at the CSO. They are listed in Table~\ref{tab:obs1} and in Table~\ref{tab:obs2} and identified with an asterisk in column 7 (named References). To do so, we have used the following assumptions:
\begin{itemize}
\item  \textbf{Henize 2-10}: we modelled the source with an axisymmetric Gaussian distribution, with a full width half maximum (FWHM) of 13'', as found by Meier et al. 2001.
\item \textbf{NGC 253}: we made use of the $^{12}$CO(6-5) map to determine the source size at 690 GHz and model the emission at higher frequencies. We modelled the nucleus as an elliptical source with Gaussian intensity profiles. The $^{12}$CO(6-5) map can be well fitted with such a model, with half maximum width of 23'' along the major axis and 11'' along the minor axis, which is very similar to the size of the CS(J=2-1) emission in the nucleus (Peng et al. 1996). The source appears to be resolved along the major axis but unresolved along the minor axis. We used the same source model for $^{12}$CO(7-6) and CI($^{3}$P$_{2}$-$^{3}$P$_{1}$), as it fits our restricted data set.
\item I \textbf{is} derived using the formula:
\vspace*{-0.5cm}
\begin{center}
\begin{equation}\label{ItoA}
\text{I} = \frac{2 \times k_{b} \times \nu^{3}}{c^{3}} \times \text{A} 
\end {equation}
\end{center}
\begin{center}
\begin{equation}\label{ItoA2}
\text{I}(Wm^{-2}sr^{-1}) = 1.02\times 10^{-18}\times \left(\frac{\nu}{GHz}\right)^{3} \times \left(\frac{A}{Kkms^{-1}}\right)
\end {equation}
\end{center}

where c is the speed of light, $\nu$ is the line frequency in $GHz$ and A is the line area in $Kkms^{-1}$.\\
To derive the flux, F in $Wm^{-2}$, we multiply I in $Wm^{-2}sr^{-1}$ by the beam solid angle $\Omega(B)$ in \textit{sr} defined as:
\vspace*{-0.5cm}
\begin{center}
\begin{equation}\label{AtoF}
  \Omega(B) = 1.133 \times \text{B}^{2} \times \frac{1}{206265^{2}} \text{ }sr
\end {equation}
\end{center}
where B is the half power beam width (HPBW) in \textit{arcsec}. The estimated error for our data in Table~\ref{tab:obs1} and in Table~\ref{tab:obs2} is about 20\%.\\
\end{itemize}

%________________________________
\subsection{C and CO cooling:}\label{secsub:cool}
Our observations are designed to provide essential information on the cooling and consequently on the thermal balance of the interstellar medium in Henize 2-10 and NGC 253. We shall also deduce which CO line(s) contribute the most to the total observed CO cooling and estimate the total observed cooling of C and CO by summing intensities in $Wm^{-2}sr^{-1}$ of all transitions listed in Table~\ref{tab:obs1} and in Table~\ref{tab:obs2} with asterisks (both literature data and our dataset). We have computed the observed C and CO cooling in the galaxy nuclei for a beam size of 21.9'', this corresponds to linear scales of 640 pc and 270 pc for Henize 2-10 and NGC 253 respectively.\\
For Henize 2-10 and NGC 253 we measured a total observed CO cooling rate of $7.2\times10^{-9}\text{ }Wm^{-2}sr^{-1}$ and $4.6\times10^{-7}\text{ }Wm^{-2}sr^{-1}$ respectively. For NGC 253 the lines contributing the most to the observed CO cooling are $^{12}$CO(6-5) (39.2\% of the total intensity) followed by $^{12}$CO(7-6) (28.3\%). For Henize 2-10, it is $^{12}$CO(7-6) (43.1\%) followed by $^{12}$CO(6-5) (31.9\%).\\
 The observed total cooling rates of neutral carbon C for Henize 2-10 and NGC 253 are respectively $3.0\times10^{-10}\text{ }Wm^{-2}sr^{-1}$ (CI($^{3}$P$_{1}$-$^{3}$P$_{0}$) transition only) and $6.7\times10^{-8}\text{ }Wm^{-2}sr^{-1}$ (for NGC 253, CI($^{3}$P$_{2}$-$^{3}$P$_{1}$) represents 52.2\% of the total).\\
These results show that $^{12}$CO(6-5) and $^{12}$CO(7-6) are contributing the most to the total observed CO cooling, with very similar percentages for both galaxies. It is natural to wonder whether $^{12}$CO(5-4) and $^{12}$CO(8-7) are strong also. This will be investigated in Sec.~\ref{secsub:lvg} and Sec.~\ref{secsub:pdr}. \\
In starburst nuclei, CO cooling is larger than C cooling by a factor of $>10$, which explains why it is easier to detect CO than C in distant galaxies. Similar results have been obtained on J1148+5251 (\textit{z}=6.42) and on  PSS2322+1944 (\textit{z}=4.12) (Bertoldi et al (2003), Walter et al. (2003) and Cox et al. (2002)). Barvainis et al (1997) detected  CI($^{3}$P$_{1}$-$^{3}$P$_{0}$) (3.6 $\pm$ 0.4 $Jykms^{-1}$) in the Cloverleaf quasar at redshift z=2.5. In the same quasar, Wei$\beta$ et al (2003) detected CI($^{3}$P$_{2}$-$^{3}$P$_{1}$) (5.2 $\pm$ 0.3 $Jykms^{-1}$) and  $^{12}$CO(3-2) (13.2 $\pm$ 0.2 $Jykms^{-1}$). In distant objects, C cooling also seems to be weaker than CO cooling.  \\

%______________________________________________________________________________
\section{CO models:}\label{sec:mod}
In this section, we use the measured CO line ratios and intensities (I and A) to determine the physical conditions of molecular gas, namely the kinetic temperature, the gas density, the CO column density and the FUV flux: $\chi$. In the first section, we start with an LTE analysis. In Sect.~\ref{secsub:lvg}, we use an LVG radiative transfer model and in Sect.~\ref{secsub:pdr}, we discuss the use of a PDR model. In the last section, we discuss the similarities and differences between the two galaxies, \textbf{the nucleus of the Milky Way and the distant QSO ``the Cloverleaf''. The latter has been chosen as being representative of distant, actively star forming galaxies.}\\

%_____________________________
\subsection{LTE analysis:}\label{secsub:lte}
In the local thermodynamical equilibrium (LTE) approximation, we
assume that CO is thermalized, hence the relative populations of its
energy levels are functions of the kinetic temperature (assumed
uniform) only. We discuss two limiting cases, 
applied to $^{12}$CO and $^{13}$CO:
\begin{itemize}
\item[$\diamond$] the lines are optically thick (optical depth: $\tau\gg1$); the line intensity ratio (eg $\frac{^{12}CO(3-2)}{^{12}CO(2-1)}$) can be written:
\begin{center}
\vspace*{-0.5cm}
\begin{equation}\label{Rthick}
  R^{thick}_{32/21} = \frac{FF \times T^{12}CO(3-2)}{FF \times T^{12}CO(2-1)} = \frac{J_{\nu32}(T_{k}) - J_{\nu32}(T_{bg})}{J_{\nu21}(T_{k}) - J_{\nu21}(T_{bg})}
\end{equation}
\end{center}
with $J_{\nu}(T_{k})$ defined as in Eq~\ref{Jnu} in Sec.~\ref{sec:intro}. In LTE, $R^{thick}_{32/21}\approx 1$ and  $R^{thick}_{76/65}\approx 1$ for $T_{k}>30\text{ }K$, while $R^{thick}_{65/32}\approx 1$ and $R^{thick}_{76/32}\approx 1$ for $T_{k}>100\text{ }K$. Plots of the line intensity ratios ($R^{thick}$) as a function of the kinetic temperature $T_{k}$ for the optically thick case in the LTE approximation are given in Fig.~\ref{fig:lte}. From this figure, we conclude that the line ratios combining high-J ($^{12}$CO(6-5) or $^{12}$CO(7-6)) with low-J ($^{12}$CO(3-2)) lines are the most useful for constraining the kinetic temperature.\\
\begin{figure}
\begin{center}
    \epsfxsize=9cm 
    \epsfbox{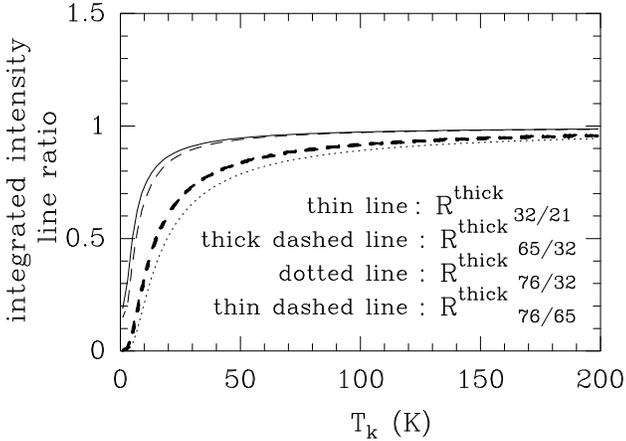} 
  \caption{Integrated intensity lines ratios, $R^{thick}$, (defined as
  in Eq.~\ref{Rthick}), vs kinetic temperature in \textit{K}. We assume LTE, and optically thick lines (optical depth, $\tau\gg1$).}\label{fig:lte}  
\end{center}
\end{figure}

\item[$\diamond$] One line is optically thick and the other is optically thin; this is likely when comparing line intensities for different isotopologues of CO in the same J transition (eg $^{12}CO(3-2)$/$^{13}CO(3-2)$). Assuming equal excitation temperatures for both species, optically thin $^{13}$CO lines and optically thick $^{12}$CO lines, we get:
\vspace*{-0.5cm}
\begin{center}
\begin{equation}\label{Rthin}
  R_{12/13} = \frac{FF \times T ^{12}CO}{FF \times T ^{13}CO} \approx \frac{A(^{12}CO)}{A(^{13}CO)} \approx \frac{1}{\tau(^{13}CO)}
\end{equation}
\end{center}
\begin{equation}\label{Rthin}
\hspace*{1.4cm}
\approx \frac{X_{galaxy}}{\tau(^{12}CO)}
\end{equation}
where $X_{galaxy}$ is the $^{12}$CO/$^{13}$CO abundance ratio. 
Previous work, using LVG models, found, for Henize 2-10, 
$X \simeq 30$ (Baas, Israel \& Koornneef 1994) and for NGC 253, $X
\simeq 40$ (Henkel \&  Mauersberger 1993; Henkel et al. 1993 and
Israel \& Baas 2002). 
From the literature data, summarized in Table~\ref{tab:obs1}, and in
Table~\ref{tab:obs2}, we deduced $\tau(^{12}$CO$(1-0)) \simeq 2.0$ and
$\tau(^{12}$CO$(2-1)) \simeq 1.6$ for Henize 2-10. For NGC 253, we
obtained $\tau(^{12}$CO$(1-0)) \simeq 3.5$ and $\tau(^{12}$CO$(2-1))
\simeq 3.1$ ($\tau$ increase for larger $X_{galaxy}$: assuming $X_{NGC
  253} \simeq 50$ as in Bradford et al. (2003) $\tau(^{12}$CO$(1-0))
\simeq 4.3$ and $\tau(^{12}$CO$(2-1)) \simeq 3.9$.

Unfortunately, the opacity of the $^{12}$CO(3-2) line for both galaxies is not accurately determined due to systematic differences between data sets (Baas, Israel \& Koornneef 1994; Harrison, Henkel \& Russell 1999 and this work). For Henize 2-10, $\tau(^{12}$CO$(3-2))$ ranges between 1.7 and 3.4, and for NGC 253, it ranges between 3.0 and 5.8 for the abundance ratio given above ($X_{NGC 253}=40$). We note that the largest error is for the $^{13}$CO integrated intensity which is the most difficult line to observe: for NGC 253, A($^{13}$CO(3-2)) varies between $90\text{ }Kkms^{-1}$ (Harrison, Henkel \& Russell 1999) and $210\text{ }Kkms^{-1}$ (Wall et al. 1991) while A($^{12}$CO(3-2)) varies between $619\text{ }Kkms^{-1}$ (Israel, White \& Baas 1995) and $1194\text{ }Kkms^{-1}$ (Wall et al. 1991) (at the same resolution).
\end{itemize}
Because the observed ratio is $R^{thick}_{65/32}\approx 0.6$ for both 
Henize 2-10 and NGC 253 (see Table~\ref{tab:rap}), we conclude, from the LTE analysis, that the $^{12}$CO(3-2) line opacities are moderate ($1<\tau<10$) and the gas is warm ($T_{k} > 30\text{ }K$). \\
To go further and constrain the gas density, we now use LVG models.\\

%____________________________
\subsection{LVG model:}\label{secsub:lvg}
The detection of the high-excitation CO lines directly indicates the
presence of large amounts of warm and dense gas in these galaxies. The
CO excitation is modelled in terms of standard, one-component, spherical
LVG radiative transfer model with uniform kinetic temperature and
density (Goldreich \& Kwan 1974; De Jong,
Dalgarno \& Chu 1975). \textbf{Though the assumption of uniform physical
conditions is crude, these models represent a useful step in
predicting intensities of the submillimeter CO lines, since, in any
case, we lack information concerning the gas distribution in the
galaxy, particularly its small scale structure.}

There are four main variables in LVG models: the CO column density
divided by the line width: N($^{12}$CO)/$\Delta v$, the molecular
hydrogen density: n(H$_{2}$), the kinetic temperature: $T_{k}$ and the
$\frac{^{12}\text{CO}}{^{13}\text{CO}}$ abundance ratio:
$X_{galaxy}$. \textbf{The linewidth, 
 $\Delta v$, is constrained by the fit of the $^{12}$CO(3-2) spectra (FWHM)}. We have used $\Delta v = 60\text{ }kms^{-1}$ for Henize 2-10 and $\Delta v = 190\text{ }kms^{-1}$ for NGC 253 (see Fig.~\ref{fig:spec}). The $\frac{^{12}\text{CO}}{^{13}\text{CO}}$ abundance ratio varies between 30 and 40 for Henize 2-10 ($X_{Henize2-10}$) and between 30 and 50 for NGC 253 ($X_{NGC253}$).\\
\textbf{To make quantitative estimates of the physical parameters of the
molecular gas in nuclei, we have compiled information about line emission for all CO rotational
transitions observed so far and scaled them to a common beam size of
21.9''(see Table~\ref{tab:obs1} and Table~\ref{tab:obs2} and
Sec.~\ref{secsub:speketmap}).}

\begin{figure*}
  \begin{center}
  \epsfxsize=14cm 
  \epsfbox{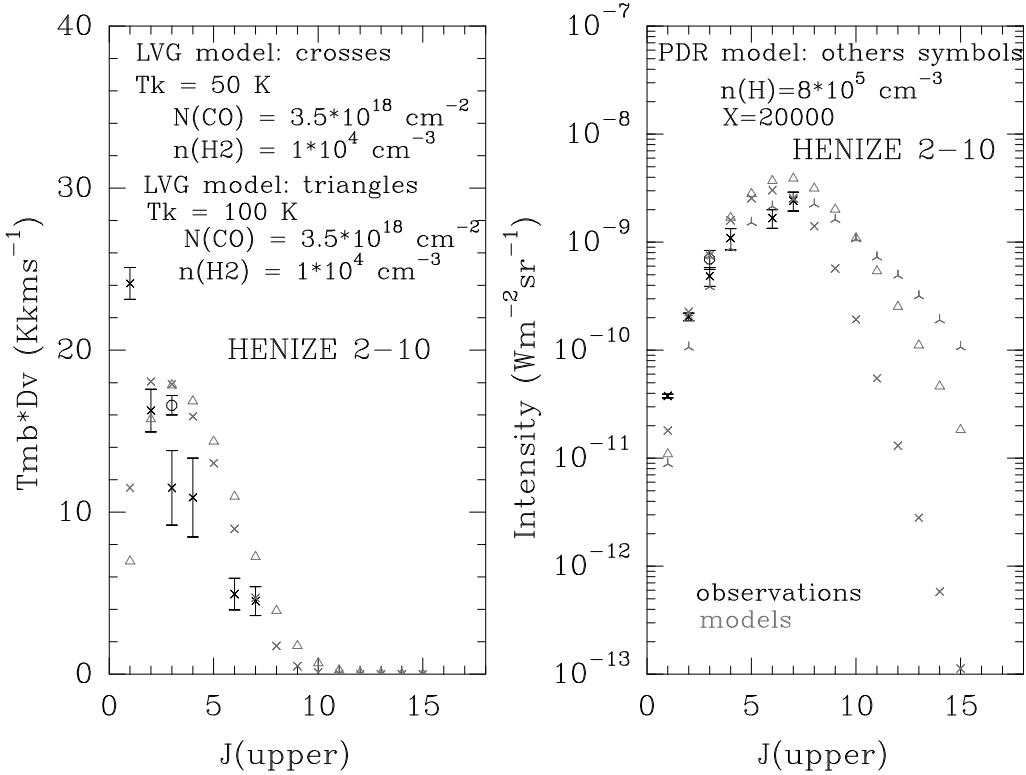} 
  \caption{LVG and PDR model calculations compared with observations
  for Henize 2-10. On the left side, we plot integrated intensities in
  $Kkms^{-1}$ vs J$_{upper}$. On the right side, we plot I in
  $Wm^{-2}sr^{-1}$ vs J$_{upper}$. In both figures, grey crosses represent
  the ``low temperature'' LVG model while grey triangles represent the
  ``high temperature'' LVG model. The starred grey triangles represent results of the PDR model (see Sect.~\ref{secsub:pdr}). Observations (with error bars) taken from literature and from our data set are shown in black. We have reported two observations for $^{12}CO(3-2)$ in Henize 2-10: the black cross is from our data and the black circle is from Meier et al. 1991. The last one seems to be agree better with the model predictions. As Henize 2-10 is a point source, a small pointing error can explain this difference.}\label{fig:modhe}
  \end{center}
\end{figure*}

\begin{figure*}
\begin{center}
    \epsfxsize=14cm 
    \epsfbox{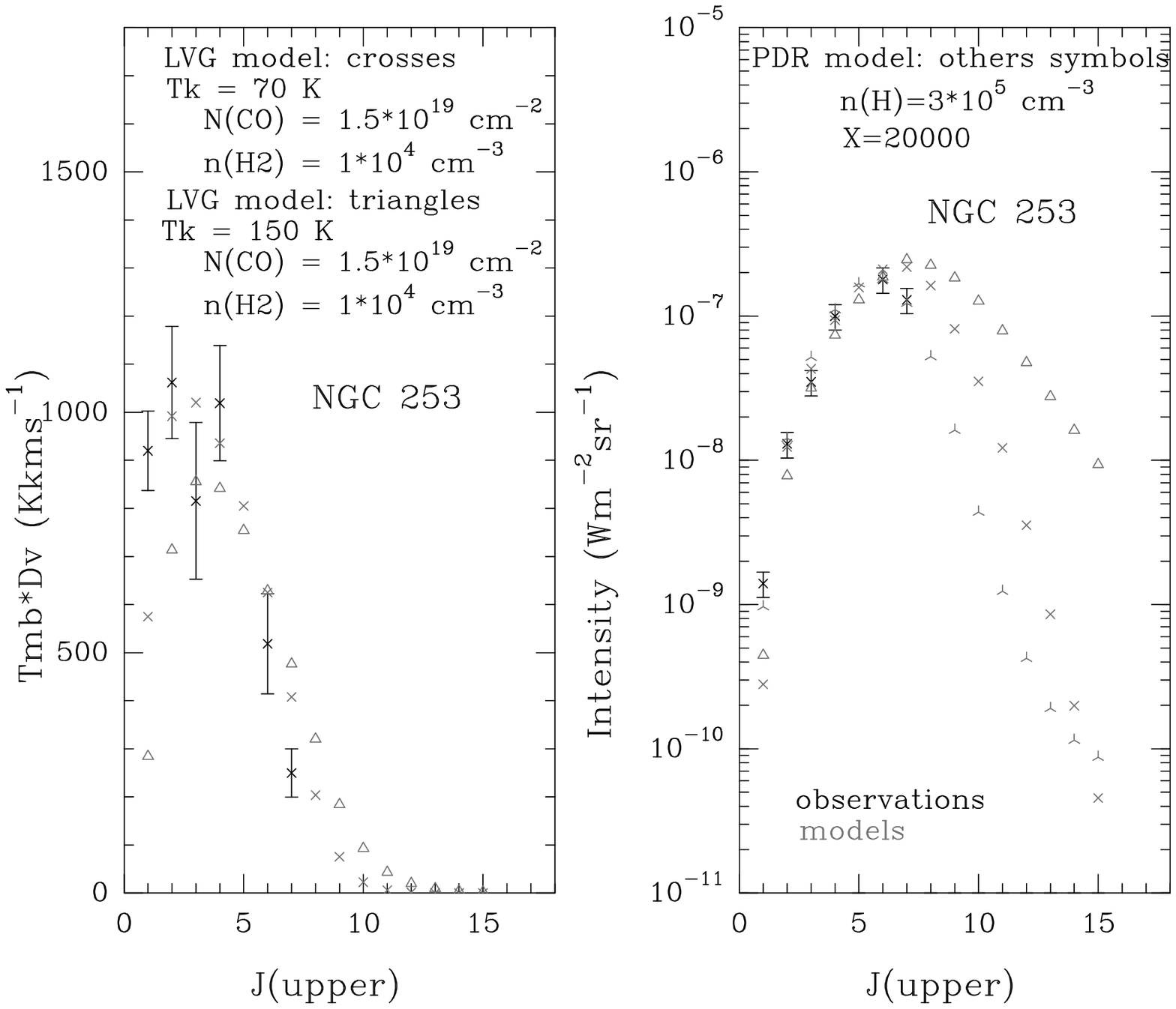} 
    \caption{LVG and PDR models calculations compared with
    observations for NGC 253. The observations are shown in black with error
    bars while model predictions are shown in grey. See caption of Fig.~\ref{fig:modhe}.}\label{fig:modng}
\end{center}
\end{figure*}

Model solutions for the physical parameters, for both sources, are presented in
Table~\ref{tab:resmod}. Because acceptable fits can be obtained over a
large domain of  parameter space, we have chosen to present two
models which bracket the range of possible solutions for the kinetic
temperature. The first one has a ``low kinetic temperature''
($T_{K}<80\text{ }K$), shown with crosses in Figs.~\ref{fig:modhe} and
~\ref{fig:modng}) and the second one has a ``high $T_{K}$''
($T_{K}>100\text{ }K$), shown with triangles in Figs.~\ref{fig:modhe}
and ~\ref{fig:modng}) (see Table ~\ref{tab:rap}). The gas density is
not well constrained but must be
at least $ 10^{4}\text{ }cm^{-3}$. The best fit $^{12}$CO/$^{13}$CO
value is 30 for Henize 2-10 and 40 for NGC 253, both compatible with
previous analyses. We varied N($^{12}$CO) from $1.0\times10^{16}\text{
}cm^{-2}$ to $1.0\times10^{20}\text{ }cm^{-2}$ for Henize 2-10 and $
3.5 \times10^{17 }\text{ }cm^{-2}$ to $ 3.5 \times10^{19 }\text{
}cm^{-2}$ for NGC 253. CO column densities are mainly constrained by
the $^{13}$CO data (see Table~\ref{tab:obs1} and
Table~\ref{tab:obs2}). Acceptable fits are found :
N($^{12}$CO)=$3.5\pm1\times10^{18}\text{ }cm^{-2}$ for Henize 2-10 and
for N($^{12}$CO)=$ 1.5 \pm0.5\times10^{19 }\text{ }cm^{-2}$ for NGC
253. We determined the ``best'' fit using a least square fit method for
the line area ratios listed in Table~\ref{tab:rap}. \textbf{As we wanted to derive the
  properties of warm gas, we have given more weight to line ratios which
  include high-J CO lines, e.g. $\frac{^{12}CO(3-2)}{^{12}CO(6-5)}$,
  $\frac{^{12}CO(3-2)}{^{12}CO(7-6)}$ and
  $\frac{^{12}CO(6-5)}{^{12}CO(7-6)}$ (see
  Sect.~\ref{secsub:lte}).}
%We used only
%ratios with high-J CO lines ($J_{upper}>5$) for this calculation, since
%we wanted to derive physical properties of the warm gas (see
%Sect.~\ref{secsub:lte}).

Predicted line integrated areas, A in $Kkms^{-1}$ and line intensities
I in $Wm^{-2}sr^{-1}$ for all CO transitions to J=15-14, are also
shown in Figs.~\ref{fig:modhe} and ~\ref{fig:modng}. The predicted
line area A is obtained by multiplying the antenna temperature, 
given by models, by the line width $\Delta v$ and by
the surface filling factor, FF. For each observed CO line, 
we estimated the surface filling
factor (FF) of molecular
clouds in the beam using the ratio $T_{mb}(observations)/T_{mb}(model)$.
\textbf{We made an  average, weighted by the S/N,  for the FF pertaining to each observed transition.} For Henize 2-10 and NGC 253, we obtained filling
factors of $FF=8.8 \pm 3 \times10^{-3}$ and  $FF=8.7 \pm 3
\times10^{-2} $ respectively. The line intensities, I, are derived
from the model A values through Eq.~\ref{ItoA} and
Eq.~\ref{ItoA2}. We give in Table~\ref{tab:rap}, line intensity ratios
of CO transitions derived from observations and from models.

\textbf{The low-J CO transitions ($^{12}$CO(1-0) and $^{12}$CO(2-1)) are not
well fit by the adopted LVG models, because we have chosen to
constrain the models with high-J transitions of CO as we are
interested in the warm gas properties. In order to study the
properties of the low excitation molecular gas, we could have
introduced another gas component in the models as in Harrison, Henkel
\& Russel (1999) or Bradford et al. (2003) for NGC 253. 
Here, the LVG models predict too low an 
intensity for the $^{12}$CO(1-0) and $^{12}$CO(2-1) lines in this
source, supporting the need for a low excitation gas component. 
For NGC 253, we obtained a similar kinetic temperature $T_{K}$ as in
Bradford et al. 2003, but a lower density $n(H_{2})$ and a higher CO
column density N($^{12}$CO). For NGC 253, Paglione et
  al. (2004) found good fits with LVG models with N($^{12}$CO)/$\Delta
  v$ between $4.0\times10^{16}\text{
  }cm^{-2}km^{-1}s$ and $1.3\times10^{18}\text{ }cm^{-2}km^{-1}s$, in
  good agreement with our value (N($^{12}$CO)/$\Delta
  v=8.0\times10^{16}\text{ }cm^{-2}km^{-1}s$).}
Recent
measurements of rotational lines of H$_{2}$ (Rigopoulou et al. 2002)
suggest that warm gas with $T_{K}\simeq 150\text{ }K$ is present in
NGC 253 which is compatible with our models.

\textbf{The LVG models have been used for predicting the CO line intensities for $^{12}$CO(1-0) up to $^{12}$CO(15-14). We have derived the total CO cooling from these predictions.}
For Henize 2-10, we obtain
with the ``low $T_{K}$'' model ($T_{K}= 50\text{ }K$),
$1.3\times10^{-8}\text{ }Wm^{-2}sr^{-1}$, and with the ``high
$T_{K}$'' model ($T_{K}= 100\text{ }K$), $2.0\times10^{-8}\text{
}Wm^{-2}sr^{-1}$. With the ``low $T_{K}$'' model, the lines which
contribute the most to the total CO cooling are $^{12}$CO(6-5)
(23.4\%), followed by $^{12}$CO(5-4) (19.7\%) and $^{12}$CO(7-6)
(19.5\%). With the ``high $T_{K}$'' model, the lines which contribute
the most to the total CO cooling are $^{12}$CO(7-6) (19.1\%) followed
by $^{12}$CO(6-5) (18.2\%) and $^{12}$CO(8-7) (15.4\%).

For NGC 253, with the ``low $T_{K}$'' model ($T_{K}= 70\text{ }K$) we
deduced a total CO cooling of $1.0\times10^{-6}\text{ }Wm^{-2}sr^{-1}$
and the most important line is $^{12}$CO(7-6) (21.1\%) followed by
$^{12}$CO(6-5) (20.4\%), $^{12}$CO(8-7) (15.7\%) and $^{12}$CO(5-4)
(15.2\%). With the ``high $T_{K}$'' model ($T_{K}= 150\text{ }K$), we
obtained a total predicted CO cooling of $1.5\times10^{-6}\text{
}Wm^{-2}sr^{-1}$, with the most intense lines being $^{12}$CO(8-7) and
$^{12}$CO(7-6) (16.4\% for each) followed by $^{12}$CO(6-5) (13.6\%)
and $^{12}$CO(9-8) (13.4\%).

$^{12}$CO(6-5) and $^{12}$CO(7-6) therefore appear to contribute
significantly to the CO cooling. Also, CO lines with $J_{upper}
\geqslant 10$ are predicted to be weak and will not have significant
antenna temperatures (see Figs.~\ref{fig:modhe} and ~\ref{fig:modng},
plots on the left side). In addition to $^{12}$CO(6-5) and
$^{12}$CO(7-6), data for $^{12}$CO(8-7) and $^{12}$CO(9-8) would be
most useful in discriminating between models, and for a more accurate
determination of the CO cooling. $^{13}$CO(6-5) data would also be
extremely useful for constraining the models and for measuring the
opacity of $^{12}$CO(6-5) line.

\begin{table}
\caption{Observed and predicted (``low $T_{K}$'' and ``high $T_{K}$'' LVG models) line area ratios for Henize 2-10 and NGC 253.}\label{tab:rap}
\begin{center}
\begin{tabular}{|c|c|c|}
\hline
 A & NGC 253 & HENIZE 2-10 \\
($Kkms^{-1}$)& observations & observations\\
\hline
  $\frac{^{12}CO(3-2)}{^{12}CO(6-5)}$ & 1.6$\pm$0.6$^{*}$  & 1.7$\pm$0.7$^{*}$\\
\hline
  $\frac{^{12}CO(3-2)}{^{12}CO(7-6)}$ & 3.3$\pm$1.3$^{*}$  & 2.0$\pm$0.8$^{*}$\\ 
\hline
 $\frac{^{12}CO(6-5)}{^{12}CO(7-6)}$ & 2.1$\pm$0.8$^{*}$  & 1.2$\pm$0.5$^{*}$\\
\hline
\hline
 $\frac{^{12}CO(1-0)}{^{13}CO(1-0)}$ & 11.5$\pm$2.2$^{*}$ & 14.9$\pm$5.8$^{*}$\\
\hline
 $\frac{^{12}CO(2-1)}{^{13}CO(2-1)}$ & 13.0$\pm$3.0$^{*}$ & 19.2$\pm$5.8$^{*}$\\
\hline
 $\frac{^{12}CO(3-2)}{^{13}CO(3-2)}$ & 9.1$\pm$3.1$^{*}$ & 8.9$\pm$3.8$^{*}$\\
\hline
\hline
& ``low $T_{K}$''  model & ``low $T_{K}$'' model\\
& $T_{K}=70\text{ }K$ & $T_{K}=50\text{ }K$\\
\hline
 $\frac{^{12}CO(3-2)}{^{12}CO(6-5)}$ & 1.6  & 2.0\\
\hline
  $\frac{^{12}CO(3-2)}{^{12}CO(7-6)}$ & 2.5  & 3.8\\
\hline
 $\frac{^{12}CO(6-5)}{^{12}CO(7-6)}$ & 1.5  & 1.9\\
\hline
\hline
 $\frac{^{12}CO(1-0)}{^{13}CO(1-0)}$ & 25.5 & 19.1\\
\hline
 $\frac{^{12}CO(2-1)}{^{13}CO(2-1)}$ & 13.4 & 9.9\\
\hline
 $\frac{^{12}CO(3-2)}{^{13}CO(3-2)}$ & 10.1 & 8.1\\
\hline
\hline
& ``high $T_{K}$'' model & ``high $T_{K}$'' model\\
& $T_{K}=150\text{ }K$ & $T_{K}=100\text{ }K$\\
\hline
 $\frac{^{12}CO(3-2)}{^{12}CO(6-5)}$ & 1.4  & 1.6\\
\hline
  $\frac{^{12}CO(3-2)}{^{12}CO(7-6)}$ & 1.8  & 2.5\\
\hline
 $\frac{^{12}CO(6-5)}{^{12}CO(7-6)}$ & 1.3  & 1.5\\
\hline
\hline
 $\frac{^{12}CO(1-0)}{^{13}CO(1-0)}$ & 15.1 & 23.9\\
\hline
 $\frac{^{12}CO(2-1)}{^{13}CO(2-1)}$ & 20.7 & 14.8\\
\hline
 $\frac{^{12}CO(3-2)}{^{13}CO(3-2)}$ & 14.8 & 9.9\\
\hline
\end{tabular}
\end{center}
$^{*}$ : ratio derived from observations marked with asterisks in Table~\ref{tab:obs1} and in Table~\ref{tab:obs2}.\\ 
\end{table}

\begin{table}
    \caption{Parameters of the ``best'' LVG models.}\label{tab:resmod}
\begin{center}
    \begin{tabular}{|c|c|c|}
      \hline
      &\textbf{Henize 2-10}& \textbf{NGC 253}\\
      \hline
      n(H$_{2}$) (cm$^{-3}$)& $\gtrsim$1 $\times$ 10$^{4}$ & $\gtrsim$1 $\times$ 10$^{4}$\\
      \hline
      N($^{12}$CO) (cm$^{-2}$) & 3.5$\pm$1 $\times$ 10$^{18}$ & 1.5$\pm$0.5 $\times$ 10$^{19}$\\
      \hline
      $\frac{^{12}C}{^{13}C}$ & 30 & 40\\
      \hline
      $\Delta v^{*}$ (kms$^{-1}$)& 60 & 187\\
      \hline
      $T_{K}$ (K)& 50-100 & 70-150\\
      \hline
      FF & 8.8$\pm$3 $\times$ 10$^{-3}$ & 8.7$\pm$3 $\times$ 10$^{-2}$\\
      \hline
    \end{tabular}
\end{center}
$^{*}$ deduced from Gaussian fits to the $^{12}CO(3-2)$ line profiles.
\end{table}


%_____________________________
\subsection{PDR model:}\label{secsub:pdr}
To progress further in the analysis of the physical conditions in the
starburst nuclei, we made use of PDR models. \textbf{Such models have
  been used here to understand the properties of the interstellar
  medium, since they take account of all the relevant physical and
  chemical processes for thermal balance. The PDR models accurately
  reproduce the steep kinetic temperature gradient near cloud edges,
  when illuminated by  intense FUV radiation.} Such models have been
developed during the past two decades, for a variety of astrophysical
sources, from giant molecular clouds illuminated by the interstellar
radiation field to the conditions experienced by circumstellar disks,
very close to hot massive stars (Tielens \& Hollenbach 1985; Van
Dishoeck \& Black 1986, 1988; Wolfire, Hollenbach \& Tielens 1990;
Hollenbach, Takahashi \& Tielens 1991; Abgrall et al. 1992; Le Bourlot
at al. 1993; K\"oster et al. 1994; Sternberg \& Dalgarno 1995; Draine
\& Bertoldi 1996, Stoerzer et al. 1996, Lhuman et al. 1997; Pak et
al. 1998; Hollenbach \& Tielens 1999 and Kaufman et al. 1999).

\textbf{We adopted here the PDR model developed by Le Bourlot et al. (1993)
for Galactic sources (see also Le Petit, Roueff \& Le Bourlot 2002) .} The source is modelled as a plane-parallel slab,
illuminated on both sides by FUV radiation to better reproduce the
starburst environment as massive stars and giant molecular clouds are
spatially correlated. Model parameters include the gas density,
assumed uniform, the intensity of the illuminating FUV radiation,
the gas phase elemental abundances, the grain properties, the gas to
dust ratio, etc.
Because the metallicities of both NGC 253 and Henize 2-10 (see
Table~\ref{tab:prop}) are close to solar, we used a model with Milky
Way abundances ($12+log(\frac{O}{H})=$8.90 $\pm$ 0.04 Boselli, Lequeux
\& Gavazzi 2002). \textbf{We have adopted standard grain properties and 
gas to dust ratio appropriate for Galactic interstellar clouds.}
We have sampled a wide range of the parameter space, varying the gas
density, n(H) and the incident FUV flux, $\chi \times G_{0}$, where
$G_{0}$ is the local average interstellar radiation field (ISRF)
determined by Draine (1978) ($G_{0}= 2.7\times 10^{-3}\text{
}ergcm^{-2}s^{-1}$). The $^{12}$C/$^{13}$C ratios ($X_{Henize2-10}$
and $X_{NGC253}$) are the same as for the ``best'' LVG  models (see
Table~\ref{tab:resmod}). Models are constrained by the ratio of integrated intensities in $ergcm^{-2}s^{-1}sr^{-1}$ listed in Table~\ref{tab:obs1} and in Table~\ref{tab:obs2} ($E$($ergcm^{-2}s^{-1}sr^{-1}$) = $I(Wm^{-2}sr^{-1})\times 10^{3}$).\\
Parameter pairs (n(H); $\chi$) which fit our dataset, are listed in
Table~\ref{tab:pdr2}. $\chi$ seems to be constrained by the
$^{12}$CO(J+1$\rightarrow$J)/$^{13}$CO(J+1$\rightarrow$J)(e.g.
$^{12}$CO(3-2)/$^{13}$CO(3-2)) ratios while n(H) is more sensitive to
the line ratios involving two $^{12}$CO lines:
$^{12}$CO(J+1$\rightarrow$J)/$^{12}$CO(J'+1$\rightarrow$J')(e.g.
$^{12}$CO(3-2)/$^{12}$CO(6-5)) ratio. The difference between
observations and model outputs is estimated to be of the order of
20\%. Emissivity ratios obtained from
the observations and from the models are listed in Table~\ref{tab:pdr}.

Model predictions for the CO line emissivities are shown in
Fig.~\ref{fig:modhe} and Fig.~\ref{fig:modng} for Henize 2-10 and NGC
253 respectively. The model predictions have been scaled to match the
observed line intensities. \textbf{As stated above, PDR models have
  been developed for local interstellar clouds. The velocity
  dispersion of the modelled cloud is a parameter in those models,
  which is used for computing the photo-dissociation rates of H$_{2}$
  and CO, and the line emissivities. This parameter is set to 1
  $kms^{-1}$, a typical figure for local molecular clouds (see
  Wolfire, Tielens \& Hollenbach 1990). Using a PDR model for fitting
  the galaxy observations is complicated by the fact that, in a
  galaxy, many PDRs contribute to the signal detected in each beam,
  resulting in a broad line (tens to hundred of kms$^{-1}$),
  compared to a single PDR line (1 kms$^{-1}$). To correct for this
effect, the model line emissivities have been multiplied by the line
width ratio, $\frac{\Delta v(galaxy)}{\Delta v(PDR)}$, where $\Delta
v(galaxy) \approx 190\text{ }kms^{-1}$ and $60\text{ }kms^{-1}$ for NGC 253 and Henize 2-10 respectively, and
$\Delta v(PDR)=1\text{ }kms^{-1}$. Once this correction is performed,
PDR model results are compared with observed data in the same way as
the LVG models (see paragraph 5 in Sect.~\ref{secsub:lvg}), and the surface
filling factor of the emission in the beam,  PDR\_FF, is computed. The
PDR\_FF is 9.4$\times 10^{-2}$ for NGC 253 and 1.5 $\times 10^{-3}$
for Henize 2-10. For NGC 253, the surface filling factors derived from
 the PDR and the LVG models are very similar. 
For Henize 2-10, the filling factor is
significantly smaller for the PDR model than for the LVG model. In
Henize 2-10, LVG models, as with PDR models, do not reproduce
observations very well due to the lack of high-J CO lines (i.e.,
$^{12}$CO(8-7) and up) which would constrain the location of the
maximum of the CO cooling curve. However, compared to LVG models, PDR
models tend to fit the series of CO lines better, particularly for NGC
253. So, we prefer the PDR values of the FF to those from the LVG.}

We have computed the total CO cooling from PDR models summing the contribution from all CO lines from $^{12}$CO(1-0) up to $^{12}$CO(15-14). In the observations, we miss $^{12}$CO(5-4) and all CO lines from $^{12}$CO(8-7) and up. For Henize 2-10, we obtained $1.5\times10^{-8}\text{ }Wm^{-2}sr^{-1}$ (similar to the CO cooling from the ``low $T_{K}$'' LVG model ($T_{K}=50\text{ }K$)). The lines which contribute the most to the total CO cooling are $^{12}$CO(7-6) (17.4\%) followed by $^{12}$CO(8-7) (15.7\%) and $^{12}$CO(6-5) (15.0\%) (with the ``low $T_{K}$'' LVG model, the lines which contribute the most are also CO lines with $J_{upper} \geqslant 5$).  For NGC 253, we deduced a total CO cooling of $7.2\times10^{-7}\text{ }Wm^{-2}sr^{-1}$, (close to, but lower than the CO cooling from the ``low $T_{K}$'' LVG model ($T_{K}=70\text{ }K$)). The most important line is $^{12}$CO(6-5) (24.9\%) followed by $^{12}$CO(5-4) (23.2\%) and $^{12}$CO(7-6) (17.0\%).\\
For NGC 253, the cooling derived from the PDR model is $\approx 40$\% higher than the observed CO cooling, while for Henize 2-10, the cooling from PDR models is a factor of 2 larger than the measured value. This can be explained by the fact that, for Henize 2-10, PDR models do not fit observations as well as for NGC 253. \textbf{Actually, $^{12}$CO(8-7) or $^{12}$CO(9-8), and $^{13}$CO(6-5) or $^{13}$CO(7-6), would be very useful in order to determine the position of the peak of the CO cooling curve, and the opacity of the $^{12}$CO(6-5) or $^{12}$CO(7-6) lines.}
%resulting in a better fit. Also, a small pointing error on this source could change a lot the value of the area (in $Kkms^{-1}$) since Henize 2-10 is a point-like object. 
Finally, the strong dependence of the PDR model fit with the density
n(H), as can be seen by comparing Fig.~\ref{fig:modhe} and
Fig.~\ref{fig:modng}, might also explain the difference between the
two galaxies. We discuss this point in the next section.

%As for the LVG models, we conclude that more data would be very useful particularly $^{12}$CO(8-7) or (9-8) and $^{13}$CO(6-5) or (7-6) in order to determine more accurately the position of the peak of the CO cooling curve and the CO column density of warm gas.\\ 

\begin{table}
    \caption{Parameters of the ``best'' PDR models.}\label{tab:pdr2}
\begin{center}
    \begin{tabular}{|c|c|c|}
      \hline
      &\textbf{Henize 2-10}& \textbf{NGC 253}\\
      \hline
      n(H) (cm$^{-3}$)& 8.0$\pm 1\times$ 10$^{5}$ & 3.0$\pm 0.5 \times$ 10$^{5}$\\
      \hline
      $\chi$ (on each side) & 20000 & 20000$^{a}$\\
      \hline
      $ v^{b}$ (kms$^{-1}$)& 60 & 187\\
      \hline
      FF & 1.5$\pm 0.5\times 10^{-3}$ & 9.7$\pm 4\times 10^{-2}$\\
      \hline
    \end{tabular}
\end{center}
$^{a}$ Carral et al. (1994) found for NGC 253 a FUV flux of 2$\times 10^{4} G_{o}$ in units of Habing (1968) corresponding to 1.2$\times 10^{4} G_{o}$ in units of Draine (1978); compatible with our value.\\
$^{b}$ deduced from Gaussian fits to line profiles.\\
\end{table} 
\begin{table}
\caption{Observed and modeled line emissivity ratios for Henize 2-10 and NGC 253 using PDR models.}\label{tab:pdr}
\begin{center}
\begin{tabular}{|c|c|c|}
\hline
 E & NGC 253 & HENIZE 2-10 \\
($ergcm^{-2}s^{-1}sr^{-1}$)& observations & observations\\
\hline
$\frac{^{12}CO(3-2)}{^{12}CO(6-5)}$ & 0.19$\pm 0.076^{*}$ & 0.22$\pm 0.083^{*}$\\
\hline
  $\frac{^{12}CO(3-2)}{^{12}CO(7-6)}$ & 0.27$\pm 0.11^{*}$ & 0.16$\pm 0.064^{*}$\\
\hline
 $\frac{^{12}CO(6-5)}{^{12}CO(7-6)}$ & 1.38$\pm 0.56^{*}$ & 0.74$\pm 0.29^{*}$\\
\hline
\hline
 $\frac{^{12}CO(1-0)}{^{13}CO(1-0)}$ & 12.73$\pm 2.5$$^{*}$ & 16.82$\pm 6.2^{*}$\\
\hline
 $\frac{^{12}CO(2-1)}{^{13}CO(2-1)}$ & 14.44$\pm 3.4$$^{*}$ & 22.22$\pm 6.1^{*}$\\
\hline
 $\frac{^{12}CO(3-2)}{^{13}CO(3-2)}$ & 10.61$\pm 3.6$$^{*}$ & 10.21$\pm 4.7^{*}$\\
\hline
\hline
& model & model\\
\hline
$\frac{^{12}CO(3-2)}{^{12}CO(6-5)}$ & 0.26 & 0.18\\
\hline
  $\frac{^{12}CO(3-2)}{^{12}CO(7-6)}$ & 0.34 & 0.15\\
\hline
 $\frac{^{12}CO(6-5)}{^{12}CO(7-6)}$ & 1.32 & 0.86\\
\hline
\hline
 $\frac{^{12}CO(1-0)}{^{13}CO(1-0)}$ & 15.42 & 15.07\\
\hline
 $\frac{^{12}CO(2-1)}{^{13}CO(2-1)}$ & 9.72 & 8.02\\
\hline
 $\frac{^{12}CO(3-2)}{^{13}CO(3-2)}$ &  7.92 & 6.27\\
\hline
 \end{tabular}
\end{center}
$^{*}$ : ratio derived from observations marked with asterisks in Table~\ref{tab:obs1} and Table~\ref{tab:obs2}.\\
\end{table} 

%____________________________________________________
\subsection{Discussion}\label{secsub:discu}
\begin{figure}
\begin{center}
    \epsfxsize=9cm 
    \epsfbox{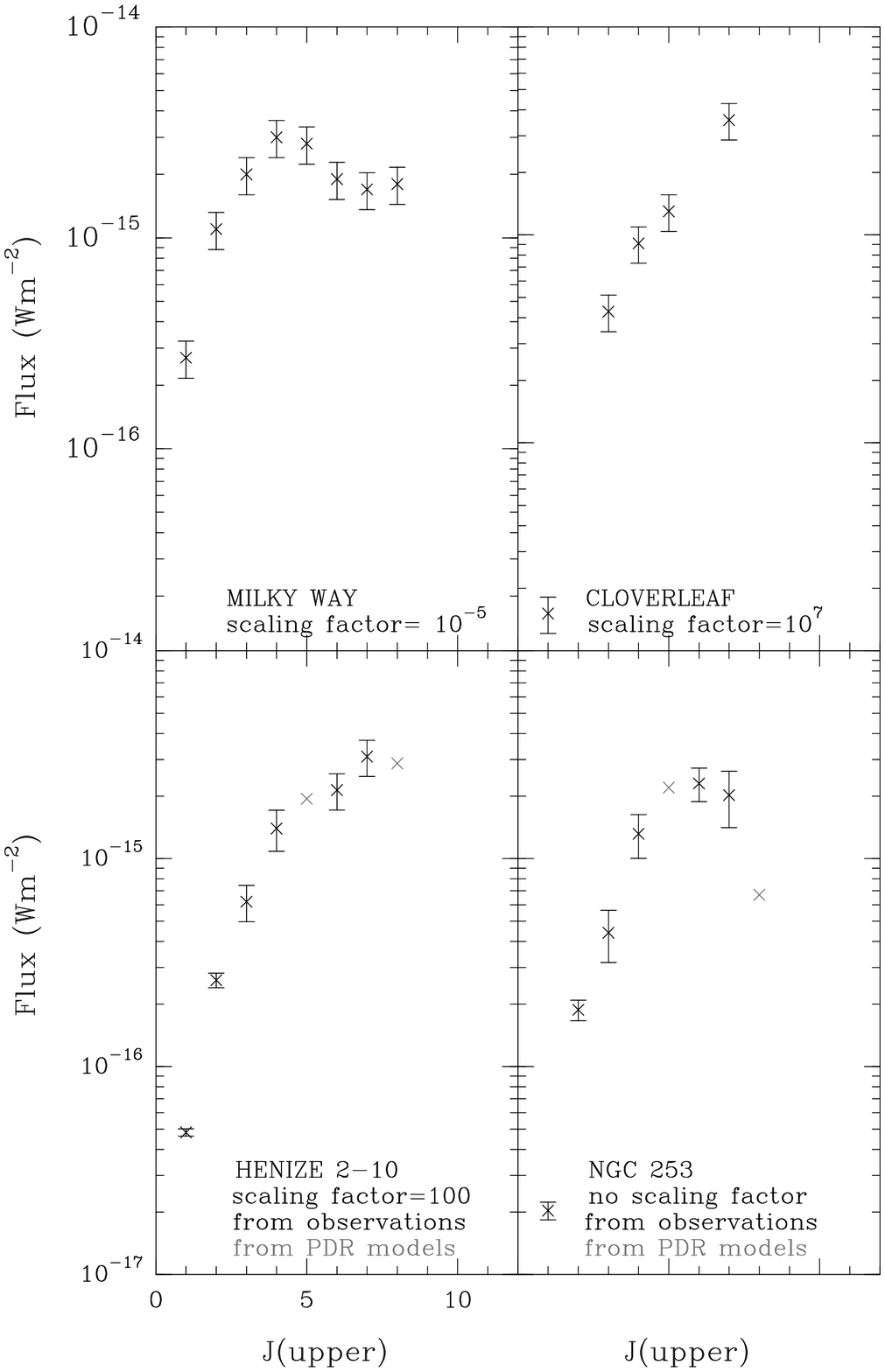} 
    \caption{Flux ($Wm^{-2}$) vs. $J_{upper}$ for the nucleus of the
    Milky Way (top left), NGC 253 (top right), Henize 2-10 (bottom
    left) and the Cloverleaf QSO (bottom right). PDR models have been
    used to obtained fluxes (plotted in grey) of $^{12}$CO(5-4) and
    $^{12}$CO(8-7) in Henize 2-10 and in NGC 253. Observations are
    shown in black. To facilitate the comparison, we applied  scaling
    factors of $1 \times 10^{7}$ for the Cloverleaf QSO data, $1 \times
    10^{-5}$ for the nucleus of the Milky Way, and 100 for Henize
    2-10. Data for NGC 253 are shown with no scaling factor.
 For the Cloverleaf QSO, the $^{12}$CO(1-0) line flux  is an upper limit (see Tsuboi et al. 1999).}\label{fig:comp}
\end{center}
\end{figure}

From the models described above, we conclude that the molecular gas in the starburst regions of Henize 2-10 and NGC 253 experience similar physical conditions: both the line ratios and derived properties are similar. Nevertheless, we can see in Fig.~\ref{fig:modhe} and in Fig.~\ref{fig:modng}, a density difference between the two galaxies. In fact, it is difficult to fit a PDR model for Henize 2-10 since we do not have a CO detection for J higher than J=7-6. The reason is \textbf{the large intensity} of the $^{12}$CO(7-6) line relative to $^{12}$CO(6-5). When the turn over in the CO cooling curve is not well constrained, fitting one PDR model is not easy and the solution can be understood as a ``lower limit''.\\
Moreover lower transitions of carbon monoxide like $^{12}$CO(1-0) and $^{12}$CO(2-1) were not really essential here since models were dedicated to the study of warm gas. So, the detection of high-J transitions was crucial for describing the starburst nuclei.\\
Another point is the influence of the FWHM line width. It is obvious for NGC 253 that the FWHM of the $^{12}$CO(3-2) line ($ \approx 190kms^{-1}$) is larger than that of the $^{12}$CO(7-6) line ($ \approx 95kms^{-1}$), the main reason being the better spatial resolution of the $^{12}$CO(7-6) data. Because NGC 253 has a steep velocity gradient along its major axis, the convolution to lower spatial resolution, using the adopted source model, is expected to produce broader lines profiles. We tried to take into account this phenomenon in our study by using LVG models with ratios in line integrated area in $Kkms^{-1}$ instead of using ratios in main beam temperature in $K$. We also studied LVG models with a smaller $\Delta v$ but this implies a lower N(CO) since the meaningful variable in LVG models is the CO column density divided by the line width: N($^{12}$CO)/$\Delta v$. So different parameter pairs (N($^{12}$CO), $\Delta v$) may reproduce the observations equally well.\\
\textbf{PDR models predict the emissivity of
the CI($^{3}$P$_{1}$-$^{3}$P$_{0}$) and the CI($^{3}$P$_{2}$-$^{3}$P$_{1}$), fine structure transitions of atomic carbon at 492 $GHz$ and 809 $GHz$ respectively
(see Table~\ref{tab:CI}).} For Henize 2-10, the predicted
CI($^{3}$P$_{1}$-$^{3}$P$_{0}$) line emissivity is brighter than the
observed value by a factor of 2. For NGC 253, the predicted
transitions are brighter than those observed by a factor of 3 for
CI($^{3}$P$_{1}$-$^{3}$P$_{0}$) and a factor of 12 for the
CI($^{3}$P$_{2}$-$^{3}$P$_{1}$). We concluded that PDR models fitted
to the CO line emission do not reproduce the atomic carbon data very
well. Perhaps, to better constrain models with observed atomic carbon
transitions, we should reduce the density. The fine-structure lines of
atomic carbon seem to share the same behaviour as the low-J CO
transitions. This tendancy is found in Galactic clouds also.\\
We compared observations of CO and [CI] in the center of the Milky Way (Fixsen et al. 1999), and in the Cloverleaf QSO (Barvainis et al. 1994, Tsuboi et al. 1999 and Wei$\beta$ et al. 2003) with our observations. \textbf{These two objects are well known and we can compare them with NGC 253 and Henize 2-10, to determine the differences or the similarities in the physical properties of their warm gas.} We chose to plot CO lines fluxes (F in $Wm^{-2}$) versus J$_{upper}$, instead of CO lines intensities (I in $Wm^{-2}sr^{-1}$) versus J$_{upper}$ as in Fig.~\ref{fig:modhe} and in Fig.~\ref{fig:modng}, because the Milky Way data found in Fixsen et al. (1999) were only available in these units. The Cloverleaf QSO data (in $Jykms^{-1}$ in Barvainis et al. 1994, Tsuboi et al. 1999 and Wei$\beta$ et al. 2003) have been converted to the same units \textbf{using} the Jy per K factors and the main beam efficiencies of telescopes (see references above). Line fluxes (F in $Wm^{-2}$) are listed in Table~\ref{tab:comp} and the flux ratios in Table~\ref{tab:rapflu}.\\ 
CO line fluxes (F in $Wm^{-2}$) for the Cloverleaf QSO, the nucleus of
the Milky Way, NGC 253 and Henize 2-10 are shown in
Fig.~\ref{fig:comp}. \textbf{To compare the CO cooling curves more easily, the
line fluxes have been scaled. The scaling factors are listed in the figure caption.}
In Fig.~\ref{fig:comp}, PDR model predictions have been used for the
CO lines which are lacking observations: $^{12}$CO(5-4) and
$^{12}$CO(8-7) for NGC 253 and for Henize 2-10 (see
Table~\ref{tab:comp}). Note that the PDR model used for Henize 2-10
PDR  does not reproduce observations as well as the PDR model used for NGC 253 (see Fig.~\ref{fig:modhe} and Fig.~\ref{fig:modng}).\\
\textbf{ We compare the CO cooling curve for the center of the
Milky Way, the Cloverleaf QSO, Henize 2-10 and NGC 253.} The turnover
positions for Henize 2-10, NGC 253 and the Cloverleaf QSO are close to
each other, near $^{12}$CO(6-5) or above, while the turnover of the CO cooling
curve of the nucleus of the Milky Way is near $^{12}$CO(4-3). We may
ask if this observed difference between the Milky Way and the three
other sources is consistent with a difference in linear resolution, or
if it is due to differences in inherent physical conditions between
the four environments. The linear resolution for NGC
253 corresponding to the adopted beam size of 21.9'' is 265 pc. For
Henize 2-10, with a distance of 6 Mpc and the same  beam size,
 we obtained a linear resolution of 637 pc. For the Cloverleaf
QSO, we computed the angular distance from $H_{0}$ and $q_{0}$ in
Wei$\beta$ et al. (2003), we obtained $D_{A}= 1056.4$ Mpc and a linear
resolution of 13 kpc. For the nucleus of the Milky Way, the 7$^{0}$
COBE-FIRAS beam gives a linear resolution of 1035 pc in the Galactic
Center (distance of 8.5 kpc). \textbf{We conclude that despite the wide
range of linear resolutions, the turnover of the CO cooling curve is found at roughly the same J value, above $^{12}$CO(6-5) for Henize 2-10, NGC 253 and the
Cloverleaf QSO. However, at the same linear scale
as Henize 2-10, the gas in the Milky Way nucleus shows less excitation.
Thus, it seems likely that differences in the shapes of the CO cooling curves
 are due to differences in the  ISM properties in
the target galaxies. The fact that the Cloverleaf QSO, NGC 253 and Henize 2-10
have the same ratios is related to the similarity of the physical
properties of warm gas for these three sources, which translates into
similar CO line ratios  (see Table~\ref{tab:rapflu})}.

\begin{table}
\caption{Results of PDR models for [CI] in Henize 2-10 and NGC 253. We
  reported also observations from Table~\ref{tab:obs1}
  and~\ref{tab:obs2}. The PDR model predictions have been multiplied 
 by the line width ratio, $\frac{\Delta v(galaxy)}{\Delta v(PDR)}$, and
  by the surface filling factor, PDR\_FF, (see Table~\ref{tab:pdr2}).}\label{tab:CI}
\begin{center}
\begin{tabular}{|c|c|c|}
\hline
 E & NGC 253 & HENIZE 2-10 \\
($ergcm^{-2}s^{-1}sr^{-1}$)& observations & observations\\
\hline
CI($^{3}$P$_{1}$-$^{3}$P$_{0}$) & 3.5$\times$ 10$^{-5}$ &  3.0$\times$ 10$^{-7}$\\
\hline
CI($^{3}$P$_{2}$-$^{3}$P$_{1}$) & 3.2$\times$ 10$^{-5}$ & not observed\\
\hline
& PDR model & PDR model\\
\hline
CI($^{3}$P$_{1}$-$^{3}$P$_{0}$) & 1.1$\times$ 10$^{-4}$ & 5.1$\times$ 10$^{-7}$\\
\hline
CI($^{3}$P$_{2}$-$^{3}$P$_{1}$) & 3.9$\times$ 10$^{-4}$ &  1.9$\times$ 10$^{-6}$\\
\hline
 \end{tabular}
\end{center}
\end{table}

\begin{table*}
\caption{Observations of the Milky Way nucleus (from Fixsen et al. 1999), NGC 253 (from this paper), Henize 2-10 (from this paper) and the Cloverleaf QSO (from Barvainis et al. 1994, Tsuboi et al. 1999 and Wei$\beta$ et al. 2003). }\label{tab:comp}
\begin{center}
\begin{tabular}{|c|c|c|c|c|}
\hline
 F  & Milky Way & NGC 253 & HENIZE 2-10 & Cloverleaf\\
($Wm^{-2}$)&&&&\\
\hline
CI($^{3}$P$_{1}$-$^{3}$P$_{0}$) &  1.9$\times$ 10$^{-10}$ & 4.6$\times$ 10$^{-16}$ &  3.8$\times$ 10$^{-18}$ & 1.7$\times$ 10$^{-23}$ \\
\hline
CI($^{3}$P$_{2}$-$^{3}$P$_{1}$)  & 1.9$\times$ 10$^{-10}$& 4.0$\times$ 10$^{-16}$ & not observed & 3.9$\times$ 10$^{-23}$\\
\hline
\hline
$^{12}$CO(1-0)  & 2.7$\times$ 10$^{-11}$& 2.0$\times$ 10$^{-17}$ & 4.8$\times$ 10$^{-19}$ & $\le$1.5$\times$ 10$^{-24}$\\
\hline
$^{12}$CO(2-1)  & 1.1$\times$ 10$^{-10}$& 1.9$\times$ 10$^{-16}$ & 2.5$\times$ 10$^{-18}$ & \\
\hline
$^{12}$CO(3-2) & 2.0$\times$ 10$^{-10}$& 4.4$\times$ 10$^{-16}$ & 9.0$\times$ 10$^{-18}$$^{a}$ & 4.3$\times$ 10$^{-23}$ \\
\hline
$^{12}$CO(4-3) & 3.0$\times$ 10$^{-10}$& 1.3$\times$ 10$^{-15}$ & 1.4$\times$ 10$^{-17}$ & 9.11$\times$ 10$^{-23}$ \\
\hline
$^{12}$CO(5-4)  & 2.8$\times$ 10$^{-10}$& 2.2$\times$ 10$^{-15}$$^{b}$ & 3.3$\times$ 10$^{-17}$$^{b}$ & 1.3$\times$ 10$^{-22}$\\
\hline
$^{12}$CO(6-5)  & 1.9$\times$ 10$^{-10}$& 2.2$\times$ 10$^{-15}$ & 2.9$\times$ 10$^{-17}$ & \\
\hline
$^{12}$CO(7-6)  & 1.7$\times$ 10$^{-10}$& 1.7$\times$ 10$^{-15}$ & 4.0$\times$ 10$^{-17}$ & 3.6$\times$ 10$^{-22}$\\
\hline
$^{12}$CO(8-7)  & 1.8$\times$ 10$^{-10}$& 6.7$\times$ 10$^{-16}$$^{b}$
 & 4.9$\times$ 10$^{-17}$$^{b}$ &  \\
\hline
 \end{tabular}
\end{center}
$^{a}$: We chose to use $^{12}$CO(3-2) observations from Meier et al. (2001) because they seem to agree with LVG and PDR models (see Fig.~\ref{fig:modhe}) better than our observations.\\
$^{b}$:These values are computed from PDR models.\\
\end{table*}

\begin{table*}
\caption{Flux ratios for the center of the Milky Way, NGC 253, Henize 2-10 and the Cloverleaf QSO.}\label{tab:rapflu}
\begin{center}
\begin{tabular}{|c|c|c|c|c|}
\hline
 F & Milky Way & NGC 253 & HENIZE 2-10 & Cloverleaf \\
($Wm^{-2}$)&&&&\\
\hline
$\frac{CI(^{3}P_{1}-^{3}P_{0})}{CI(^{3}P_{2}-^{3}P_{1})}$  & 1.00 & 1.15 & not observed & 0.42\\
\hline
$\frac{^{12}CO(3-2)}{^{12}CO(1-0)}$ & 7.37 & 22 & 18.75 & $\ge$ 28.6\\
\hline
 $\frac{^{12}CO(3-2)}{^{12}CO(2-1)}$ & 1.84 & 2.32 & 3.6 & - \\
\hline
$\frac{^{12}CO(3-2)}{^{12}CO(4-3)}$  & 0.67 & 0.34 & 0.64 & 0.47\\
\hline
 $\frac{^{12}CO(3-2)}{^{12}CO(5-4)}$  & 0.72 & 0.21$^{*}$ & 0.27$^{*}$ & 0.33\\
\hline
 $\frac{^{12}CO(3-2)}{^{12}CO(6-5)}$  & 1.03 & 0.20 & 0.31 & - \\
\hline
$\frac{^{12}CO(3-2)}{^{12}CO(7-6)}$ & 1.18 & 0.26 & 0.23 & 0.12\\
\hline
$\frac{^{12}CO(3-2)}{^{12}CO(8-7)}$ & 1.09 & 0.66$^{*}$ & 0.18$^{*}$ & - \\
\hline
 \end{tabular}
\end{center}
$^{*}$:These values are computed from PDR models.\\
\end{table*}

\section{Conclusions:}
We observed Henize 2-10 in the rotational lines of carbon monoxide $^{12}$CO(J=3-2), (J=4-3), (J=6-5) and (J=7-6) and NGC 253 in the $^{3}$P$_{2}$-$^{3}$P$_{1}$ fine structure transitions of atomic carbon [CI] at 809 GHz and in the  rotational lines of carbon monoxide $^{12}$CO(3-2), (J=6-5) and (J=7-6). We show that C cooling is less than CO cooling for both galaxy nuclei by a factor $>$ 10. Among observed lines, those which contribute most to the CO cooling are $^{12}$CO(6-5) and $^{12}$CO(7-6).\\
Such high-J transitions are needed to constrain the physical
conditions in starburst nuclei. We used both LVG and PDR models for
each galaxy: the molecular gas in the Henize 2-10 nucleus is well described
with a LVG model defined by $\Delta V$ = 60 kms$^{-1}$, $T_{k}\approx
50-100\text{ }K$, $\frac{^{12}CO}{^{13}CO} \approx 30$, $n(H_{2})
\gtrsim 10^4\text{ }cm^{-3}$ and
$N(^{12}CO)=3.5\pm1\times10^{18}\text{ }cm^{-2}$. For the NGC 253
nucleus, we derived the following parameters: $\Delta V$ = 190
kms$^{-1}$, $T_{k}\approx 70-150\text{ }K$, $\frac{^{12}CO}{^{13}CO} \approx 40$, $n(H_{2}) \gtrsim 10^4\text{ }cm^{-3}$ and $N(^{12}CO)=1.5 \pm 0.5\times10^{19}\text{ }cm^{-2}$. PDR models provide equally good fits to the data. PDR models were used to give us more information about the physical parameters in these media. We succeeded in reproducing the observations with an accuracy of about 20\% using a PDR model defined for Henize 2-10 by n(H)=$8.0\pm 1 \times 10^{5}$ $cm^{-3}$ and $\chi=20000$ (we modelled the source as a plane-parallel slab, illuminated on both sides by FUV radiation). For NGC 253, we derived the following model: n(H)=$3.0\pm 0.5 \times 10^{5}$ $cm^{-3}$ and $\chi=20000$.\\
Thanks to those models, we predict that, for distant galaxies, to
obtain properties of cold and warm gas, the most interesting CO lines to be observed are $^{12}$CO(1-0), (2-1), (3-2), (4-3), (5-4), (6-5), (7-6) and $^{12}$CO(8-7). $^{12}$CO(9-8) and higher-J CO lines will be weaker and more difficult to detect in distant galaxies. The reasons are twofold. First the most intense lines are $^{12}$CO(6-5) and $^{12}$CO(7-6), second, though $^{12}$CO(9-8) can have a significant contribution to the cooling, its antenna temperature is significantly lower than for other CO lines (eg $^{12}$CO(4-3) or $^{12}$CO(6-5)). Also, the typical $T_{K}$ and n($H_{2}$) values only allow significant excitation over a large fraction of the galaxy for J$_{up}\leqslant$10 transitions. For higher-J transitions the values of $T_{K}$ and n($H_{2}$) high enough to excite J$_{up}\geqslant$10 will be confined to a volume too small to produce a detectable signal.\\
 Data on the $^{12}$CO(8-7) and on the $^{13}$CO(6-5) lines
   would be extremely useful in further studies; $^{12}$CO(8-7) will
   help to localize the maximum of the CO cooling curve, while
   $^{13}$CO(6-5) will help constrain the warm gas column density. Even now,
   it is still difficult to detect these lines from the ground; with
   APEX receivers which may cover atmospheric windows up to a frequency of
   1.4 THz \footnote{http://www.mpifr-bonn.mpg.de/div/mm/apex.html},
   these lines could become available. \textbf{ALMA  will give access  to
   better spatial resolution, for resolving individual molecular
   clouds :} 
%``\textit{Baselines as large as 3 km at 230 GHz are definitely warranted in order to study high-mass star-forming cores at large distances at the same linear scale as Orion-KL. The rms sensitivity of 6-12 K in a 1 MHz bandwidth in 1 hr integration at 0.08" resolution is sufficient to detect many of the stronger lines, which typically have T$_{b}>$ 20 K}''. 
\textbf{And with HIFI and PACS on board the Herschel satellite, lines with frequencies up to 5 THz will become accessible.}\\
We compared properties of warm gas derived from our models with the properties of the nucleus of the Milky Way and of the Cloverleaf QSO. We concluded that the ISM in NGC 253, Henize 2-10 and in the Cloverleaf QSO are similar, leading to similar CO excitation, while the nucleus of the Milky Way exhibits lower excitation CO lines.\\  
%We have estimated the total mass of molecular gas, M(H$_{2}$), and the fraction of warm gas in both galaxies. We obtained for Henize 2-10: $M(H_{2})= 4.8\pm 20\text{\%}\times 10 ^{7} M_{\odot}$ and $M_{W}(H_{2})=3.0 \pm 0.7 \times 10 ^{6} M_{\odot}$ depending on the LVG model used. For NGC 253 we estimated $M(H_{2})=6.1\pm 20\text{\%}\times 10 ^{7} M_{\odot}$ and $M_{W}(H_{2})=9.3 \pm 3.2 \times 10 ^{7} M_{\odot}$. Those values are consistent with other results from the literature. The warm gas amounts to $\lesssim$ 15\% of the total gas mass but contributes most of the CO cooling.\\

\begin{acknowledgements} 
This work has benefitted from financial support from CNRS-PCMI and CNRS-INSU travel grants.
We thank J. Cernicharo for letting us use his CO LVG model and J. Le Bourlot et P. Hily-Blant for introducing us to his PDR model.
The CSO is funded by the NSF under contract \# AST 9980846. We thanks the referee for the useful comments.
\end{acknowledgements}

\end{document}